\documentclass[12pt]{article}
\usepackage{hyperref}
\usepackage{cite}
\usepackage{color}
\usepackage{graphicx}
\usepackage{amsmath}
\usepackage{amssymb}
\usepackage{xspace}

\makeatletter
\@addtoreset{equation}{section}

\makeatletter
\renewcommand\section{\@startsection {section}{1}{\z@}%
                                   {-3.5ex \@plus -1ex \@minus -.2ex}
                                   {2.3ex \@plus.2ex}%
                                   {\normalfont\large\bfseries}}
\renewcommand\subsection{\@startsection{subsection}{2}{\z@}%
                                     {-3.25ex\@plus -1ex \@minus -.2ex}%
                                     {1.5ex \@plus .2ex}%
                                     {\normalfont\bfseries}}

\def\baselinestretch{1.2}
\newcommand{\be}{\begin{equation}}
\newcommand{\ee}{\end{equation}}
\newcommand{\beq}{\begin{eqnarray}}
\newcommand{\eeq}{\end{eqnarray}}

\newcommand{\gone}[1]{{}}

\begin{document}
\begin{titlepage}
\begin{flushright}
MAD-TH-13-02
\end{flushright}

\vfil

\begin{center}

{\bf \Large Gravity dual of dynamically broken supersymmetry}

\vfil

William Cottrell, J\'er\^ome Gaillard, and Akikazu Hashimoto

\vfil

Department of Physics, University of Wisconsin, Madison, WI
53706, USA

\vfil

\end{center}

\begin{abstract}
\noindent We study a renormalization group flow of ABJM theory
embedded into the warped $A_8$ geometry and explore the dependence of
the vacuum structure on the parameters of the theory. This model has a
product group gauge structure $U(N)_k \times U(n+l)_{-k}$ and comes
equipped with discrete parameters $N$, $l$, and $k$, a continuous
parameter $b_\infty$ related to the ratio of the Yang-Mills coupling
for the two gauge groups, and one dimensionful parameter $g_{YM}^2$
setting the overall scale. A supersymmetric supergravity solution
exists when $Q = N - l(l-k)/2k - k/24$ is positive and is
interpretable as a RG flow from a Yang-Mills like UV fixed point to a
superconformal IR fixed point with free energy of order $Q^{3/2}$. The
fate of the theory when $Q$ is taken to be negative is less clear. We
explore the structure of the possible gravity solution for small
negative $Q$ by considering the linearized gravitational back reaction
from adding a small number of anti-branes on the $Q=0$ background.
Following the work of Bena, Gra\~na, and Halmagyi, we find that a sensible
solution satisfying appropriate boundary conditions does not appear to
exist. This leaves the status of the RG flow for the $Q<0$ theories a
mystery. We offer the following speculative resolution to the puzzle:
the $-k/24$ unit of charge induced by the curvature correction to
supergravity should be considered an allowed physical object, and one should be adding an anti brane not to the $Q=0$ background but
rather the $Q = -k/24$ background.  Such a solution has a
repulson singularity, and gives rise to a picture of the vacuum
configuration where a cluster of anti-branes are floating around the
repulson singularity, but are stabilized from being pushed off to
infinity by other fluxes. Such a state is non-supersymmetric and
appears to describe a vacuum with dynamical breaking of
supersymmetry. Based on these considerations, we construct a phase
diagram for this theory exhibiting various interesting regions.
\end{abstract}
\vspace{0.5in}

\end{titlepage}
\renewcommand{\baselinestretch}{1.05}  

\section{Introduction}
\label{sec1}

Dynamical supersymmetry breaking (DSB) is an important phenomenon in
supersymmetric field theories. It is a critical ingredient in model
building where one aims to incorporate both the benefits of
supersymmetry and the empirical fact that this symmetry is not always
manifest. Several concrete models exhibiting DSB are known and are
reviewed, for example, in \cite{Shadmi:1999jy}. It would be
interesting, nonetheless, to extend the list of examples of models
exhibiting DSB and to study this phenomenon from new perspectives such
as gauge/gravity correspondence.

The model of Aharony, Bergman, Jafferis, and Maldacena (ABJM) is a
promising framework to explore this issue. As a field theory, this
model is a Chern-Simons theory with product gauge group and level
$U(N)_k \times U(N+l)_{-k}$ with specific matter content and
interactions \cite{Aharony:2008ug,Aharony:2008gk}. For $k > 1$ this
model has ${\cal N}=6$ superconformal symmetry. Its candidate gravity
dual is M-theory on $AdS_4 \times S^7/Z_k$ with $l$ units of discrete
torsion supported by the orbifold. The free energy of this theory can
be inferred from the standard Bekenstein-Hawking analysis and was
found to exhibit the familiar scaling
\cite{Bergman:2009zh,Aharony:2009fc,Drukker:2010nc}
\be F  = -{2 \sqrt{2} \over 3} k^2  \left(Q \over k\right)^{3/2}, \label{sugraF} \ee
where 
\be Q = N - {l(l-k) \over 2k} - {k \over 24} \ . \label{Qdef} \ee
This precise form for $Q$ takes into consideration the Freed-Witten anomaly as well as the contribution from the curvature \cite{Aharony:2009fc}. The expression (\ref{sugraF})   based on supergravity is the leading approximation in the planar limit $k \gg 1$ as well as in the limit of large 't Hooft coupling
\be \lambda = {Q \over k} \ . \ee
In the very remarkable work of \cite{Drukker:2010nc}, the exact
$\lambda$ dependence of the free energy in the leading planar
approximation, reproducing the leading large $\lambda$ dependence
(\ref{sugraF}) including the $-k/24$ curvature contribution, was
computed on the field theory side using the localization technique of
\cite{Kapustin:2009kz}.  This is a highly non-trivial test of the
gauge gravity correspondence for the ABJM system.

One very interesting feature of (\ref{sugraF}) is its non-analyticity
at $Q=0$. It is certainly possible to find a combination of integer
parameters $N$, $l$, and $k$ to make $Q <0$, but for such a value of
$Q$, the free energy ceases to be real. This should be interpreted as
an indication that the assumption of superconformal symmetry of the
field theory which goes into the localization analysis is breaking
down when $Q$ becomes negative. Does this mean that a field theory
with these sets of parameters is intrinsically ill behaved, or does it
simply mean that the theory exists in some gapped phase? It is not
immediately clear how to address this issue short of solving the
theory completely.

One can attempt to study this issue from the perspective of the dual
gravity formulation. However, the candidate dual geometry of $AdS_4
\times S^7/Z_k$ with radius $Q$ ceases to exist when $Q$ becomes
negative, and it is not clear how one should proceed dealing directly
with the gauge gravity duality in the ABJM theory.

One way in which we can gain some perspective on this issue is to
consider embedding the ABJM theory in some renormalization group flow
where the ultra-violet theory is expected to be well behaved for
general values of $N$, $l$, and $k$. One can then study the phase of
this theory as we vary these parameters. If the ultra-violet theory
also admits a gravity dual, one can explore the phase structure of
this theory from the structure of the full gravity solution.

Perhaps the most natural ultra-violet embedding of the ABJM theory is
to turn on a Yang-Mills term for each of the gauge groups $U(N)$ and
$U(N+l)$. The structure of the gravity dual of this embedding is
reasonably well understood
\cite{Aharony:2009fc,Hashimoto:2008iv}. Unfortunately, the gravity
dual is more complicated in structure than that of a simple
cohomogeneity one solutions, making it extremely cumbersome to work
with this setup.

Fortunately, there is another ultra-violet embedding of the ABJM
theory which is simpler on the gravity side\cite{Hashimoto:2010bq},
based on M-theory on an eight-dimensional $spin(7)$ holonomy manifold of
cohomogeneity one, known as the $A_8$ geometry, originally constructed
by Cveti\v{c}, Gibbons, Lu, and Pope \cite{Cvetic:2001pga}. Roughly
speaking, the $A_8$ manifold interpolates between $R^7 \times S^1$ at
infinity, to $R^8$ in the core, with a $U(1)$ symmetry corresponding
to translation in the $S^1$ at infinity. Near the core, this shift
rotates the $R^8$ around the origin in such a way that taking the
$Z_k$ orbifold of this $U(1)$ symmetry gives rise to a $R^8/Z_k$
orbifold precisely of the kind which appears in the construction of the
ABJM model. In order to completely formulate the gravity dual, we need
to know the self-dual 4-form on the eight dimensional $A_8$ geometry
and the warp factor sourced by charges and fluxes. All of these
structures have been constructed explicitly in \cite{Cvetic:2001pga}
and can be utilized in interpreting these solutions from the point of
view of a holographic renormalization group flow \cite{Hashimoto:2010bq}.

We will primarily work with $k \gg 1$ corresponding to the planar limit. In this limit, the radius of the asymptotic $S^1$  is small, and it is best to view the supergravity solution using the language of type IIA supergravity.

In order to explore the fate of $Q<0$ theory in this framework, it is
helpful to understand the theory with $Q=0$, which corresponds to
suitably selecting $N$, $l$, and $k$. From the gravity dual
perspective, shifting from $Q=0$ to $Q>0$ corresponds, roughly, to
adding D2-branes. Along similar lines, one expects that shifting from
$Q=0$ to $Q<0$ would correspond to adding anti D2-branes to the $Q=0$
solution. So the problem of understanding the gravity dual of the
$Q<0$ theory appears to involve the study of the gravitational back
reaction of anti-D2 branes in a certain warped geometry with fluxes.

In general, this is a cumbersome problem where one must deal with the
effects of the branes on fluxes and vice versa which can be
complicated. To make the problem more tractable on the first pass, it
is convenient to treat the shift away from $Q=0$ as being small and to
work to first order in that perturbation.

In fact, a very similar problem, computing the gravitational
back reaction of anti D3-branes to first order around the
Klebanov-Strassler background \cite{Klebanov:2000hb}, was initiated in
the work of Bena, Gra\~na, and Halmagyi \cite{Bena:2009xk} and has been
followed up in many papers including
\cite{Bena:2010gs,Bena:2011hz,Bena:2011wh,Giecold:2011gw,Bena:2012bk}.\footnote{An earlier attempt to study these backgrounds can be found in
\cite{DeWolfe:2008zy}.} Much of the analysis for the $A_8$ geometry is
similar to these works and so we will largely follow their template. It
should also be noted that despite serious efforts on the part of these
authors, these papers report on the {\it non-existence} of the back
reacted solution which is consistent with the expected boundary
conditions.

Here in this article, we will re-examine the parallel issue in the
$A_8$ setting.  There are a few salient features of the $A_8$ setup
which are distinct from the earlier works which we should point out.
\begin{enumerate}

\item One of the main motivations for considering the back reaction of
anti-branes in the Klebanov-Strassler geometry was to study the candidate
for a gravity dual of a meta-stable vacuum. As we will review shortly,
the non-supersymmetric configuration in the case of $A_8$ is expected
to correspond to the true vacuum with dynamical supersymmetry
breaking. Unlike the meta-stable vacuum which can afford not to exist
in a strongly interacting theory, one expects the true vacuum to exist
assuming that the field theory exists.

\item Unlike the 3+1 dimensional construction of Klebanov-Strassler, a
2+1 dimensional construction has a simpler UV structure. Instead of
the indefinite cascade, the theory crosses over into a
super-renormalizable 2+1 dimensional ultra-violet fixed point. This
reduces the risk of causing confusion when identifying the parameters
of the gravity solution and the field theory in the ultra-violet
region. This advantage of 2+1 dimensions was also highlighted in
\cite{Bena:2010gs,Giecold:2011gw}.

\item The warped deformed conifold asymptotes near the core to a
deformed cone $R^3 \times S^3$ and as such the anti-D3 sources were
smeared along the $S^3$ to simplify the gravity problem.  It is
generally believed that this is not a serious problem. Nonetheless,
the fact that all the earlier works fail to find the candidate
gravitational back reaction have caused many to question if this
smearing is in part the culprit. In the case of $A_8$, there will
be no room for such a doubt since the geometry of the $Q=0$ solution
has a unique origin where the anti-D2 will naturally sit and preserve
the cohomogeneity one structure without any smearing.
\end{enumerate}

Despite all these differences, the conclusion of our perturbative
analysis follows the trend set by our predecessors. We, too, find that
there are no solution interpretable as the linearized perturbation
around $Q=0$ to describe the $Q<0$ background.

In light of this finding, we will offer our speculation on how one
should think about the ultimate fate of the $Q<0$ theory.

The organization of this paper is as follows. We will start in section
2 by reviewing the warped $A_8$ geometry and its interpretation as a
gravity dual to a ultra-violet embedding of the ABJM model. We will
then review the perturbative analysis around the $Q=0$ solution
following the template of \cite{Bena:2009xk} in section 3. We will
offer our conclusions and speculations in section 4.

\section{Review of the warped $A_8$ background}

In this section, we review the warped $A_8$ background. We will begin
by reviewing the supergravity solution in subsection \ref{sgA8}. In
subsection \ref{ftA8}, we will review the dynamics of the field theory
which can be inferred from the brane description of the theory.

\subsection{Supergravity solution\label{sgA8}}

In this subsection, we will recall the essential features of the
warped $A_8$ geometry originally constructed by
\cite{Cvetic:2001pga}. Most of what we review can be found in section
4 of \cite{Hashimoto:2010bq}.
We start by considering an eight dimensional $spin(7)$ holonomy manifold
\be ds_{A_8}^2 = h(r)^2 dr^2  + \ell^2 (a(r)^2  (D \mu^i)^2  + b(r)^2 \sigma^2 + c(r)^2d \Omega_4) \label{a8ansatz}\ee
where $\sigma^2$ and $(D \mu^i)^2$ are line elements of the $S^3$
fiber on an $S^4$ base where $S^3$ itself is viewed as a $S^1$ fiber
over an $S^2$ base \cite{Cvetic:2001pga}. Functions $a(r)$, $b(r)$,
$c(r)$, and $h(r)$ are given by
\beq h(r)^2 & = & {(r+\ell)^2 \over (r+3 \ell)(r - \ell)} \cr
a(r)^2 & = & {1 \over 4\ell^2}(r+3 \ell) (r - \ell) \cr
b(r)^2 & = & {(r + 3 \ell)(r - \ell) \over (r+\ell)^2} \cr
c(r)^2 & = & {1 \over 2 \ell^2} (r^2 -\ell^2) \ .  \label{habc}
\eeq
The parameter $\ell$ sets the scale of this geometry. Topologically, this space is $R^8$. Geometrically, for large $r$, this geometry has the structure of $R^7 \times S^1$. We will consider orbifolding the coordinate $\varphi$ so that it is periodic under $4 \pi/k$. The fixed $r$ slice of this geometry has the topology of a squashed $S^7/Z_k$ which can also be viewed as a $U(1)$ bundle over a squashed $CP^3$.

The self-dual 4-form on this geometry is also known explicitly. It is
given by
\be G_4 = d C_3 \ee
where
\be C_3 =   m B_{(3)}+w d \sigma \wedge d \varphi \label{alphaeq} \ee
and
\be B_{(3)} =   v_1(r) \sigma \wedge X_{(2)} + v_2(r) \sigma \wedge Y_{(2)}+ v_3(r) Y_{(3)} \ee
with
\beq v_1(r) & = & -{(r-\ell)^2 \over 8 (r + \ell)^2} \cr
v_2(r) & = & {(r-\ell)^2(r + 5 \ell) \over 8 (r+\ell)(r + 3 \ell)^2} \\
v_3(r) & = &  - {(r-\ell)^2 \over 16 (r+3 \ell)^2} 
\eeq
as is given in \cite{Cvetic:2001pga}. We have also included a locally
exact term proportional to $w$ which turns out to play an important
role in quantizing the charges.\footnote{In \cite{Hashimoto:2010bq},
$w$ was referred to as $\alpha$.} $m$ is an adjustable parameter for
the time being.

We consider embedding this eight dimensional space in M-theory. The
resulting geometry will have 2+1 Poincar\'e symmetry. The self-dual
4-form embeds naturally as components of the M-theory 4-form field
strength. It sources the M-theory 4-form
electrically through the equation
\be d *_{11} F_4 = {1 \over 2} F_4 \wedge F_4 + (2 \pi l^{11}_p)^6  k Q \delta^8(\vec r) \,  .\label{gwedgeg}\ee
where we included the possibility of additional charge source parametrized by $Q$.  All of the M-theory equations can be solved by the ansatz
\beq ds^2 &=& H^{-2/3} (-dt^2+dx_1^2 + dx_2^2) + H^{1/3} ds_{A_8}^2 \cr
F_4 & = & dt \wedge dx_1 \wedge dx_2 \wedge d  H^{-1} +  G_{4}
\eeq
with
\beq H(r) &=& {24 \pi^2 (l^{11}_p)^6 k Q \over \ell^6} H_1(r)+{m^2 \over \ell^6} H_2(r) \label{warpH} \\
H_1(r) & = & 
\frac{\ell(3 r^3-3 r^2
   \ell -11 r \ell^2+ 27 \ell^3)}{192  (r-\ell)^3 (3
   \ell+r)} + 
 \frac{1}{256}\log \left(\frac{r-\ell}{r+3 \ell}\right)\\
H_2(r) & = & \frac{\ell^5 \left(63 \ell^2+26 r \ell+3
   r^2\right)}{20 (r+\ell)^2 (r+3 \ell)^5}
\eeq
We have anticipated taking the $\alpha'\rightarrow 0$ decoupling limit
and dropped the ``1'' term in $H(r)$. So as $r \rightarrow \infty$, $H
\rightarrow 0$.  For $k \gg 1$, it is convenient to work in the type
IIA description by reducing along the $\varphi$ coordinate. In the IIA
reduction, the M-theory 3-form $C_3$ gives rise to a IIA NSNS $B_2$
field given by
\be B_2 = {2 \over k R}  m (r-\ell)^2 \left(-{1 \over 8 (r+\ell)^2} X_{(2)} + {(r+5\ell) \over 8 (r+\ell) (r+3\ell)^2} Y_{(2)} \right) + {2 \over k R} w (X_{(2)} - Y_{(2)}) \ . \ee
where 
\be R = g_s l_s = {2 \ell \over k} \ee
is the radius of the M-theory circle and
\be l_p^{11} = g_s^{1/3} l_s \ . \ee

In order to identify the proper quantization of parameters $Q$, $m$, and $w$ which we have introduced in this solution, we need to compute the D2 and the D4 Page charge following  \cite{Hashimoto:2010bq}. This gives rise to
\beq Q &=& N_2 - {l(l-k) \over 2k} - {k \over 24} \cr
m & = & -(4 \pi g_s l_s^3) M \cr
(2 \pi)^2 w  & = & - (2 \pi l_s)^3 g_s \left(l - {k \over 2}\right)
\eeq
where
\be M = l - {k \over 2} + b_\infty k \ee
and
\be b_\infty = {1 \over (2 \pi l_s)^2} \int_{CP^1} B (r = \infty) \ee
is the period of the NSNS 2-form $B$ through the $CP^1$ cycle of the $CP^3$ at $r = \infty$ and is one of the parameters of the background. 

With this quantization condition, we find that the D2 Maxwell charge is given by
\be Q_2^{Maxwell} =  \left. {1 \over g_s (2 \pi l_s)^3} \int_{CP^3} r^6 (-H'(r))\right|_{r=\infty} = Q + {M^2 \over 2k} \ee
$Q$, on the other hand, is immediately interpretable as the brane charge. 

In order to take the $l_s \rightarrow 0$ limit, we scale $g_s$ and $r$ keeping
\be U= {r \over l_s^2}, \qquad g_{YM}^2 = {g_s \over l_s} \ee
fixed as usual.   $b_\infty$ is interpreted, as usual, as providing the gauge coupling
\be {1 \over g_{YM1}^2} = {b_\infty \over g_{YM}^2}, \qquad 
{1 \over g_{YM2}^2} = {(1-b_\infty) \over g_{YM}^2} \label{gaugecoupling} \ee

We have therefore arrived at a family of supergravity solutions,
parametrized by $N$, $l$, and $k$ which are discrete dimensionless
parameters, $b_\infty$ which is a continuous dimensionless parameter,
and $g_{YM}^2$ being the one dimensionful parameter setting the scale
of the problem. For fixed $k$ and $b_\infty$, one can parametrize the
remaining choice of models in terms of a set of parameters $(Q,M)$ in
lieu of $(N,l)$.

Note that $Q$ and $M$ are invariant under the transformation
\be  N_2 \rightarrow N_2 + l, \qquad l \rightarrow l+ k , \qquad b_\infty \rightarrow b_\infty -1 \ . \label{largegauge} \ee
This is the manifestation of large gauge transformation outlined in
\cite{Aharony:2009fc} and the invariance of $Q$ and $M$ is the
result of gauge invariance of physical quantities such as the
brane charge and the Maxwell charge.

\subsection{Properties of the warped $A_8$ solution}

In this subsection, let us review some of the basic features of the warped $A_8$ solution.

The solution is parametrized by three discrete parameters $N$, $l$,
and $k$, a continuous parameter $b_\infty$, and a scale $g_{YM}^2$.  It
is convenient to work with some fixed $b_\infty$. We will also take
$k$ to be large so that the planar approximation is reliable. Finally,
we will scale
\be N = x k, \qquad l = y k \ee
so that in the large $k$ limit, $x$ and $y$ can be viewed as continuous parameters washing out the granularity of the integers $N$ and $l$. In terms of $x$ and $y$, we can express
\be {Q \over k} = x - {y (y-1) \over 2} - {1 \over 24}, \qquad {M \over k} = y - {1 \over 2} + b_\infty \ . \ee
The Maxwell charge can then be expressed as
\be {Q_2^{Maxwell} \over k} = {Q \over k} + {M^2 \over 2 k^2}  = x + \left(y-{1 \over 2}\right) b_\infty + {1 \over 2 } b_\infty + {1 \over 12} \ . \ee
This is the quantity which we need be large in order suppress the $\alpha'$ corrections at least for a large part of the bulk region  in the gravity dual. 

The physical characteristic of this supergravity background depends sensitively on the sign of $Q$. 

For positive $Q$, the solution asymptotes to $AdS_4 \times S^7/Z_k$ in
the deep IR region. This is the dual gravity description of the ABJM
superconformal fixed point. As one flows up the holographic
renormalization group flow, the period\footnote{Note that the period
$b(r)$ is not the same as the function $b(r)$ in the ansatz
(\ref{a8ansatz}) for the $A_8$ geometry. Hopefully, it is clear from
the context which $b(r)$ we are talking about.}
\be b(r) = {1 \over (2 \pi l_s)^2} \int_{CP^1} B(r) \ee
runs from 
\be b(r=\ell) = -{l \over k} + {1 \over 2} \ee
to
\be b(r=\infty) = b_\infty \ . \ee
Each time $b(r)$ goes outside the range $0<b(r) < 1$, one can apply
the large gauge transformation (\ref{largegauge}) to bring it back
into that range. The resulting change in $N$ and $l$ is the
manifestation of the duality cascade. Unlike the case of the deformed
conifold where the cascade continues forever, here $b(r)$ approaches a
limiting value $b_\infty$ after undergoing finitely many cascades.  The
formula for the gauge coupling (\ref{gaugecoupling}) makes sense only in
the gauge where $0 < b_\infty < 1$.

Consider now the case where $Q=0$, while keeping
$Q_2^{Maxwell} > 0$. The structure of the solution is not so
dramatically changed in the ultra-violet region. However, in the
infra-red region, the solution looks very different. The term
proportional to $Q$ in the warp factor (\ref{warpH}) is gone, and
$H(r)$ approaches a finite value as $r$ approaches $\ell$. This means
that the geometry is regular at $r=\ell$. This is suggestive of the
field theory exhibiting a mass gap. One simple way to holographically
estimate the mass gap is to compute the time, in field theory
coordinate, for a light signal to travel from the boundary to $r =
\ell$ \cite{Horowitz:1999gf}
\be t_{gap} = \int_\ell^\infty {dt \over dr} dr = \int_\ell^\infty \sqrt{- {g_{rr} \over g_{tt}}} dr 
= \int_\ell^\infty H^{1/2}(r) h(r) dr \sim {|M| \over g_{YM}^2 k^2}
\ee
from which we can read off the scale
\be E_{gap} \sim {1 \over t_{gap}} \sim {g_{YM}^2 k^2 \over |M|} \sim {g_{YM}^2 k^{3/2} \over N^{1/2}} \label{Egap}\ee
where in the last relation, we used $N \gg k$, $Q=0$, and assumed $b_\infty$ is of order one.

Finally, consider what happens when $Q$ is taken to be
negative. Now, we see that $H_1(r)$ and $H_2(r)$ contribute to $H(r)$
with opposite signs in (\ref{warpH}). Since $H_1(r)$ diverges at
$r=\ell$, we learn that the background exhibits a naked singularity when
$H(r)$ becomes zero at some value of $r > \ell$. Presumably, this
singularity stems from extrapolating the background supported by
positive charge, positive tension BPS sources to negative charge and
negative tension. Since negative tension objects are unphysical, what
one must do to continue beyond the $Q=0$ background to negative $Q$ is
to add a positive tension negative charge object, i.e. an anti
D2-brane.

We have therefore arrived at a conclusion that in order to explore the
physics of $Q<0$ model, we must consider the gravitational back
reaction from adding an anti D2-brane to the $Q=0$ background. In this
sense, the problem is very similar to the program of
\cite{Bena:2009xk}.

\subsection{Field theory dynamics\label{ftA8}}

Before proceeding to analyze the gravitational back reaction of anti
D2-branes let us review our expectation from the field theory
considerations.

One disadvantage of the ultra-violet embedding based on $A_8$ as
opposed to turning on the Yang-Mills coupling for the ABJM theory is
the fact that the field theory dual of the $A_8$ construction is not
known at the same level of detail. For example, the precise form of
the Lagrangian defining the field theory dual has not been written
down. Nonetheless, one can infer quite a lot from the form of the
background on the gravity side, as well as from the consideration of
the associated brane construction.

The warped $A_8$ supergravity solution asymptotes to a squashed $CP^3$
cone in type IIA theory warped by $Q_2^{Maxwell}$ units of electric
four form flux through $CP^3$ with $b_\infty$ unit of $B_2$ on the
unique $CP^1$ cycle of the $CP^3$. So the dual field theory appears to
resemble a Yang-Mills theory with a product gauge group which is
superrenormalizable, and the gravity description is taking over as the
effective description below the energy of order \cite{Itzhaki:1998dd}
\be E = g_{YM}^2 Q_2^{Maxwell} \ .\ee

The supergravity solution also suggests that the theory flows to ABJM
in the infra red. This suggests, just as in the case of the ABJM
theory, that this model can be engineered as a decoupling limit of a
Hanany-Witten like construction of branes stretched between
overlapping 5-branes separated along a compact dimension
\cite{Aharony:2008ug,Aharony:2008gk,Aharony:2009fc}. We also know from
the structure of the supergravity solution that this background
preserves ${\cal N}=1$ supersymmetry in the 2+1 dimensional sense
\cite{Cvetic:2001pga}. A catalog of overlapping 5-brane configuration,
which we will refer to as the KOO table, was presented by Kitao, Ohta,
and Ohta in Table 1 of \cite{Kitao:1998mf}.  In the classification of the
KOO table, the ABJM construction appears to correspond to the item
4-(iii). In contrast, the natural candidate dual of the $A_8$
construction is 4-(i).

The brane construction provides a natural interpretation of the
parameter $b_\infty$ as well as the structure of the large gauge
transformation (\ref{largegauge}). The $b_\infty$ parametrizes the
distance between the 5-branes, which runs as a result of the brane
bending effect \cite{Witten:1997sc} but asymptotes to a fixed value at
large separation. The integer $N$ can be interpreted as the number of
the integer D3-branes winding all the way around the compact direction
separating the 5-branes, and $l$ is the number of fractional D3-branes
stretching between the D5-branes. The large gauge transformation
(\ref{largegauge}) can then be seen as corresponding to the way in
which $N$, $l$, and $b_\infty$ transform as one gradually slides
$b_\infty$ by one, causing one of the 5-branes to circumnavigate the
compact direction, undergoing Hanany-Witten transitions when the two
5-branes cross. These phenomena are reviewed in
\cite{Aharony:2009fc,Hashimoto:2010bq}.

\begin{figure}[t]
\centerline{\includegraphics{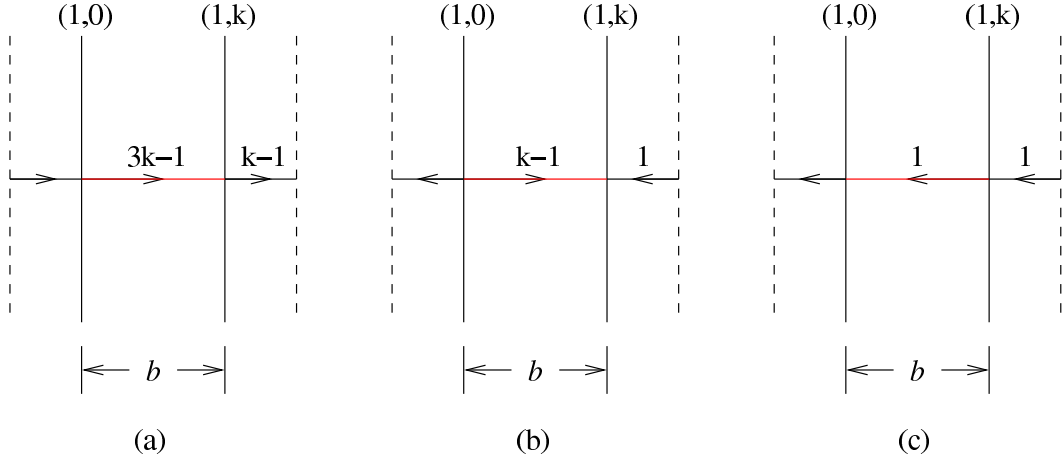}}
\caption{Hanany-Witten brane diagram for configurations violating the
generalized $s$-rule. The configurations (a), (b), and (c) are related
by sliding the $(1,k)$ brane around the circle. In this figure, labels
such as ``$3k-1$'' and ``$k-1$'' refers to the number of D3 brane
segments stretched between the 5-branes, as opposed to the counting of
integer and fractional branes. The configuration (a) corresponds to
$N=k-1$ and $l = 2k$. Configuration (b) corresponds to $N=-1$ and $l =
k$. (c) corresponds to $N=-1$ and $l=0$. This figure originally appeared in \cite{Hashimoto:2010bq}.
\label{figa}}
\end{figure}

In the framework of brane construction, it is relatively easy to see
the difference between the cases when $Q/k \gg 0$ and $Q/k \ll 0$.
For $Q/k \gg 0$, $N$ will transform in $n$ cascade steps to
\be N \rightarrow N + n l + {n(n-1) \over 2} k \ge Q - {k \over 12} \ee 
and so as long as $Q/k > 1/12$, then $N$ is positive definite. But as
$Q$ gets smaller, we will encounter a duality cascade where $N$ can
become negative. An example where this happens is illustrated in
figure \ref{figa}.

\begin{figure}[t]
\centerline{\includegraphics{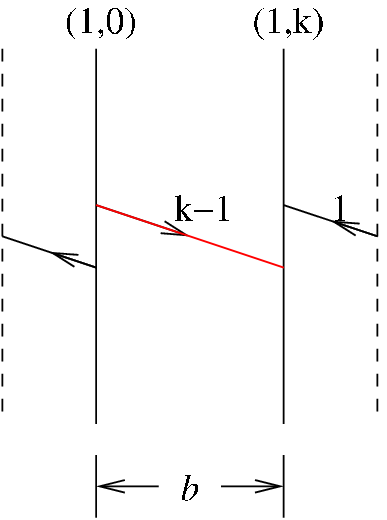}}
\caption{Schematic sketch of the expected minimum energy configuration
for the construction illustrated in figure \ref{figa}.b including the
effect of  repulsion between the brane segments. This figure originally appeared in \cite{Hashimoto:2010bq}.  \label{figb}}
\end{figure}

What has been previously noted in \cite{Hashimoto:2008iv} is that the
configuration, illustrated in figure \ref{figa}.b, corresponds to a
non-BPS stable configuration found in \cite{Mukhi:2000dn} by balancing
the repulsive force experienced by the D3 segments and the attractive
force arising from the angle of the 5-branes forcing the D3-brane to get
longer as they move apart. Schematically, one expects the stable,
non-BPS configuration to look like what is illustrated in figure
\ref{figb}.

This brane configuration suggests that the vacuum configuration breaks
supersymmetry. Strictly speaking, one should view this claim as being
valid only for the brane theory and whether or not this feature
survives the field theory limit $\alpha' \rightarrow 0$ needs to be
examined closely. One of the goals of this paper is to examine this
issue for the decoupled theory by looking at the gravity dual.

Much of what we described so far is very similar to the construction
of metastable vacua which was the motivation of the work of
\cite{Bena:2009xk} for the case of the deformed conifold.  In that
case, the decay channel to the supersymmetric ground state has
been identified \cite{Kachru:2002gs} and the stable supersymmetric
vacuum to which the system decays is easy to construct.

In the case of the $A_8$ background, no comparable decay mechanism or alternate supersymmetric geometry with the same Page charges appear
to exist. Therefore, if we were to find the non-supersymmetric
supergravity solution with the appropriate charges, it is natural to
interpret it as the gravity dual of a vacuum having undergone
dynamical supersymmetry breaking.

\section{Linearized analysis of non-supersymmetric perturbations \label{sec:lin}}

In this section, we will describe the analysis of linearized
perturbations around the $Q=0$ background. Our goal is to identify the
linearized perturbation corresponding to adding a small number of anti
D2-branes to the $Q=0$ background. A very similar problem has been
analyzed in \cite{Bena:2010gs} and \cite{Giecold:2011gw}. Reference
\cite{Giecold:2011gw} in fact considers a closely related background
also by Cveti\v{c}, Gibbons, Lu, and Pope \cite{Cvetic:2001ma}. It is
therefore extremely convenient to follow the template of the analysis
of \cite{Giecold:2011gw} for our background. The main difference
between $A_8$ and the background of \cite{Cvetic:2001ma} is that the
former asymptotes to $R^8$ whereas the later asymptotes to $R^5 \times
S^3$ in the core region. Also, the former, in the IIA description,
includes the D6 charge $k$ in addition to the integer D2 charge $N$
and the fractional D2 (integer D4) charge $l$. The holonomy and the
number of supersymmetries are also slightly different.

Just as was the case in the previous studies, 
\cite{Bena:2009xk,Bena:2010gs,Bena:2011hz,Bena:2011wh,Giecold:2011gw,Bena:2012bk}
it is convenient to employ the method of Borokhov and Gubser
\cite{Borokhov:2002fm} to look for the non-supersymmetric linearized
perturbation, corresponding to adding a small number of anti
D2-branes, to the BPS background at $Q=0$. It turns out that there is
one subtlety which manifests itself in the presence of the D2, D4, and
D6 charges which requires special attention. The issue stems from
distinguishing between D2 charges generated by explicitly adding a D2
brane from the D2 charge induced by gradually increasing the NSNS
$B$-field in the presence of a D4-brane inducing an effective D2-brane
charge.  The former changes the brane charge and the Page charge
without changing $b_\infty$. The latter changes the brane charge and
$b_\infty$ but does not change the Page charge. Strictly speaking,
from the point of view of classical supergravity which is only
sensitive to $H_3= dB_2$, the two procedures are indistinguishable. Yet,
they are distinct in the full quantum interpretation and in the
context of gauge/gravity duality.

The issue of varying a locally exact piece of NSNS 2-form with a period on some
2-cycle in the presence of a $p+2$ brane wrapping that 2-cycle is a bit
subtle when accounting for the charge of a $p$ brane as was reviewed
in appendix A of \cite{Aharony:2009fc}. At the level of classical
supergravity, one can imagine deforming the background by a) adding a locally
exact 2-form to the $B$-field without adding any additional charge
source, b) add a charge source without changing the $B$-field
asymptotically, or c) perform a combination of the two. 

Clearly, only two out of these three deformations are linearly independent
at the level of classical gravity. If one does not distinguish backgrounds
which differ only by a closed term in $B_2$ so that the two $H_3$ are
indistinguishable, only one out of these three deformations would
appear to be physical. In the $A_8$, however, $b_\infty$ is a
parameter specifying the background that must be kept account of. One
must therefore be very explicit in making sure both of the linearly
independent components of a), b), and c) are included in the space
of deformations.  This issue is closely related to the fact that we
have D2, D4, and D6 charges, and that tuning the locally exact part of $B$ affects
not only the D2 brane charge but also the D4 brane charge, whereas we
wish to adjust them independently.

In order to spell out this issue, we find it convenient to first
study the linearized supergravity analysis for the anti D2-branes in
flat space. This will turn out to also be a useful framework to review
the formalism Borokhov and Gubser. After working out this simple
exercise of identifying the anti D2-brane in flat space, it is
straightforward to generalize the procedure to the warped $A_8$ case
and to highlight the important features.

\subsection{Linearized analysis for branes in flat space \label{sec:flat}}

Consider the truncated M-theory action
\be 
S=\int d^{11}x\sqrt{-\det(g_{11})}\left(R-\frac{1}{2}\left|F_{4}\right|^{2}\right)
\ee
and consider a simple ansatz
\beq
ds^{2} & = & e^{-2z}\eta_{\mu\upsilon}dx^{\mu}dx^{\upsilon}+e^{z}\left(h^{2}dr^{2}+\ell^{2} g^{2}d\Omega_{7}^{2}\right)\label{eq:Metric Ansatz}\\
C_{3} & = & e^{-3\tilde{z}}dx^{0}\wedge dx^{1}\wedge dx^{2}\nonumber 
\eeq
where $z$, $\tilde z$, $g$, and $h$ are the a priori independent fields. To allow for the possibility of finding non-BPS solutions, we are parametrizing the warp factor
\be H = e^{3z} \ee
and the electric 3 form potential
\be \tilde H^{-1} = e^{-3 \tilde z} \ee
as independent variables. The effective action for the radial dependence of these fields takes the form
\be
S_{eff}= \int dr \, g^{7}h\left(42 \frac{1}{g^{2}}+42\frac{\ell^2}{h^{2}}\left(\frac{g'}{g}\right)^{2}-\frac{9\ell^2}{2h^{2}}(z')^{2}+\frac{9 \ell^2}{2h^{2}}(\tilde{z}')^{2}e^{6\left(z-\tilde{z}\right)}\right)\label{eq:Radial Lag 1}
\ee
The field $h$ is non-dynamical and reflects the fact that it can be
fixed to take on an arbitrary form by reparametrizing the radial
variable. Let us choose\footnote{We are treating $\tau$ as a
dimensionless variable whereas $r$ has the dimension of length.}
\be {1 \over \ell} h dr = - g^7 d\tau  \ee
so that the effective action becomes
\be
S=\int d\tau\ \left(42g^{12}+42\left(\frac{g'}{g}\right)^{2}-\frac{9}{2}\left(z'\right)^{2}+\frac{9}{2}\left(\tilde{z}'\right)^{2}e^{6\left(z-\tilde{z}\right)}\right) \label{effS}
\ee
The solutions derived from this effective action are also subject to
the zero energy condition
\be 
\left(42g^{12}-42\left(\frac{g'}{g}\right)^{2}+\frac{9}{2}\left(z'\right)^{2}-\frac{9}{2}\left(\tilde{z}'\right)^{2}e^{6\left(z-\tilde{z}\right)}\right) = 0 \label{zeroenergycond}
\ee
from varying (\ref{eq:Radial Lag 1}) with respect to $h$. 

In \cite{Giecold:2011gw}, a trick is used to substitute
\be K = - (e^{-3 \tilde z})' \label{Kdef} \ee
and write the action in the form
\be
S=\int d\tau\ \left(42g^{12}+42\left(\frac{g'}{g}\right)^{2}-\frac{9}{2}\left(z'\right)^{2}+\frac{1}{2} K^2 e^{6z}\right)
\ee
and eliminate $K$ algebraically. Such introduction of auxiliary
variable $K$ is useful for later purposes when we set up a
superpotential to characterize the BPS equations and their small
perturbations, but is not quite correct in the present form.  The
equation of motion derived from variation of $\tilde z$ implies that
$K$ is constant, not zero.

One way to address this is to include a Lagrange multiplier field
$q(\tau)$ to impose the constraint that $K = 3 \tilde z' e^{-3 \tilde
z}$
\be
S=\int d\tau\ \left(42g^{12}+42\left(\frac{g'}{g}\right)^{2}-\frac{9}{2}\left(z'\right)^{2}+\frac{1}{2} K^2 e^{6z} + q(\tau) (K + (e^{-3\tilde z})') \right)
\ee
Then, integrating out $q$ and then $K$ will reproduce
(\ref{effS}). If, instead, one integrates out $\tilde z$ first, one
infers that $q(t)=q=\mbox{constant}$. Further integrating out $K$
takes the effective action to the form
\be
S=\int d\tau\ \left(42g^{12}+42\left(\frac{g'}{g}\right)^{2}-\frac{9}{2}\left(z'\right)^{2}-\frac{1}{2} q^2 e^{-6z}\right) \label{effS2}
\ee
The parameter $q$ enters as one of the variables controlling
\be-(e^{-3 \tilde z})' = K = q e^{-6z} \label{ztK} \ee 
and integrating this equation will give rise to one more integration
constant, associated with the degrees of freedom $\tilde z$.

We can now proceed to analyze the BPS background and a first order
deformation around it following the method of \cite{Borokhov:2002fm}.

First, note that the effective action (\ref{effS2}) can be written in
the form
\be \int d\tau \, (T - U) \ee
where 
\be
U=-\frac{1}{2}G^{ij}\frac{\partial W}{\partial\phi^{i}}\frac{\partial W}{\partial\phi^{j}}
\ee
\be T = {1 \over 2} G_{ij} (\phi^i)'(\phi^j)'
\ee
for
\be \{\phi^1 , \phi^2 \} = \{g,z\}\ee
\be G_{ij}  =  \left(\begin{array}{cc} {84 \over g^2} &  \\ & -9 \end{array}\right)_{ij} \ee
and 
\be W = -14 g^6 + q e^{-3z} \ee

A BPS solution can be found by solving
\be {d \phi^i \over  d\tau}  - G^{ij} {\partial  W \over \partial \phi^j} = 0  \ . \label{BPSeq}
\ee
The solution is 
\beq \phi_0^1 & = & (6 \tau)^{-1/6} \\
\phi_0^2 & = &   {1 \over 3} \log (q \tau + 1)
\eeq
which translates to
\beq e^{3z} &=& 1 + {q \ell^6 \over 6 r^6}  \cr
g & = & {r \over \ell} \label{phi0}
\eeq
under 
\be r^6 = \ell^6 (6 \tau)^{-1} \ . \ee
Here, $\ell$ is some generic length scale introduced to keep track of
dimensions. In order to normalize $q$ to the standard M-theory charge
conventions, we see that we should set
\be {q \ell^6 \over 6} = 32 \pi^2 (l^{11}_p)^6 Q \ . \label{qNorm} \ee

To study the solution to the equation of motion at first order, we expand
\be \phi^i= \phi^i_0 + \phi^i_1 \ee
and derive the equation
\beq
\frac{d\xi_{i}}{d\tau} & = & -\xi_{j}N^{j}{}_i , 
\qquad
N^i{}_j = {\partial \over \partial \phi^j} G^{ik} {\partial W \over \partial \phi^k} 
\label{BG}\\
\frac{d\phi_{1}^{i}}{d\tau} & = & N^i{}_j\phi_{1}^{j}+G^{ij}\xi_{j}\ , \nonumber  
\eeq
satisfied by the $\phi^i_1$, as well as the set of auxiliary field
$\xi_i$. In the Borokhov-Gubser formalism, $n$ sets of second order
differential equations for the $\phi^i_1$ fields are reformulated as
$2n$ sets of first order differential equations for the $\phi_i$'s and
the $\xi_i$'s.

These equations are solved while treating $q$ also as a first order
perturbation.  However, since $q$ is not one of the parameters which
affects the fields $\xi_i$ and $\phi_1^i$ at the linear order, the
term linear in $q$ needs to be included as part of the zero-th order
solution.  We are therefore considering an expansion around the zeroth
order solution
\beq \phi_0^1 & = & (6 \tau)^{-1/6}\cr
\phi_0^2 & = &  {1 \over 3} q \tau
\eeq
and these are the background functions which go into $N_i^j$ and $G^{ij}$ in  (\ref{BG})

These equations are solved in the following order.
\begin{enumerate}
\item $\xi_2$ is dictated by
\be {d \xi_2 \over d \tau} =  0 \ee
and is solved by
\be \xi_2 = X_2 \ee
\item The equation for $\phi_1^2$ is given by
\be {d \phi_1^2 \over d \tau} =  - {1 \over 9} X_2 \ee
and is solved by
\be \phi_1^2 =   -{1 \over 9} X_2 \tau + Y_2 \ee
\item The equation for $\xi_1$ is given by
\be {d \xi_1 \over d\tau} = {7 \over 6 \tau}   \xi_1 \ee
and is solved by
\be \xi_1  =  X_1 (6\tau)^{7 /6} \ee
\item Finally, $\phi_1^1$ satisfies
\be
\frac{d\phi_{1}^{1}}{d\tau}=-\frac{7}{6\tau}\phi_{1}^{1}+\frac{1}{84}X_{1}(6\tau)^{5/6}
\ee
and is solved by
\be \phi_1^1 = {X_1 \over 1521} (6\tau)^{11 / 6} + Y_1 (6\tau)^{-7/6} \ee
\end{enumerate}

The zero energy condition (\ref{zeroenergycond}) for these fields is given by
\be \xi_i {d \phi_0^i \over d \tau} = - X_1 = 0  \ee

The general linearized solution we found can be summarized as
\beq g & = & 
(6 \tau)^{-1/6} + \left(Y_1 (6\tau)^{-7/6} \right)\\
H & = & e^{3z} = 1 +   \left( q \tau  -{1 \over 3} X_2 \tau + Y_2 \right) \\
\tilde H^{-1} & = & e^{-3 \tilde z} = - q \tau + \delta
\eeq
where $\delta$ is the integration constant we inherit from integrating
(\ref{ztK}).  In terms of the more conventional radial variable
\be r^6 = \ell^6 (6 \tau)^{-1} \ee
these solutions take the form
\beq g & = & 
r + \left( Y_1 {r^7 \over \ell^7}  \right) \\
H & = & e^{3z} = 1 +   \left( \left(q -{1 \over 3} X_2\right){\ell^6 \over r^6}  + Y_2 \right) \\
\tilde H^{-1} & = &  e^{-3 \tilde z} = -q {\ell^6 \over r^6} + \delta
\eeq
Special cases of these expressions correspond to familiar solutions. It
is convenient to set $Y_1=Y_2$ so that the geometry asymptotes to flat
space in the canonical metric for large $r$. The parameter $\delta$ is
pure gauge and so it can be set to zero without any harm. Further setting
$X_2=0$ will give rise to the linearized form of the BPS solution
identified earlier (\ref{phi0}) for $q>0$ \cite{Horowitz:1991cd} with
$q$ normalized according to (\ref{qNorm}). Choosing $X_2 =6q$ and
taking $q<0$ corresponds to what we would identify as the anti
M2-brane. Other generic values of $X_2$ appear to correspond to the
solution found in (5.12) of \cite{Gregory:1995qh} with $\alpha=0$,
$D=11$, $p=2$. One can also continue to compute higher order
corrections to $\phi_2^i$, $\phi_3^i$, etc, and reproduce the entire
non-linear solutions \cite{Horowitz:1991cd,Gregory:1995qh}.

\subsection{Back reaction of anti-branes in warped $A_8$ geometry}

We will now face the beast, and address the problem of computing the
gravitational back reaction of anti D2-branes for the warped $A_8$
background working to first order around the $Q=0$ background.

The supergravity equations of motion are inferred from the full bosonic M-theory action
\be
S={1 \over 2\kappa^{2}} \int d^{11}x\, \sqrt{-\det(g_{11})}\left(R-\frac{1}{2}\left|F_{4}\right|^{2}\right)
-\frac{1}{6}\int A_{3}\wedge F_{4}\wedge F_{4}
\label{eq:Action}
\ee
The ansatz we consider is a generalization of what we reviewed in
section \ref{sgA8} where the warp factor $e^{3z}$ and the electric
components of the 4-form $e^{3 \tilde z}$ are allowed to vary
independently. Explicitly, they are given by
\begin{eqnarray}
ds_{11}^{2} & = & e^{-2z}\eta_{\mu\upsilon}dx^{\mu}dx^{\upsilon}+e^{z}\left(ds_{8}^{2}\right)\nonumber \\
ds_{8}^{2} & = & h^2 dr^{2}+\ell^{2}\left(a^{2}\left(D\mu^{i}\right)^{2}+b^{2}\sigma^{2}+c^{2}d\Omega_{4}^{2}\right) \label{ansatz} \\
C_{3} & = & e^{-3\tilde{z}}dx^{0}\wedge dx^{1}\wedge dx^{2}+ m\left(v_{1}\sigma\wedge X_{(2)}+v_{2}\sigma\wedge Y_{(2)}+v_{3}Y_{(3)}\right)+w\, d\sigma\wedge d \varphi \nonumber 
\end{eqnarray}
We are looking for a solution which depends on a single radial variable $r$. When this ansatz is substituted  into the action (\ref{eq:Action}), we can derive a lengthy effective action
\beq S & = & \int d r \, \left\{ \frac{a^{2}bc^{4}}{h}\biggl[2(\alpha')^{2}+12(\gamma')^{2}+4\alpha'\beta'+16\alpha'\gamma'+8\beta'\gamma'-\frac{9}{2}(z')^{2}+\frac{9}{2}(\tilde{z}')^{2}e^{6(z-\tilde{z})}\biggl]\right. \nonumber \\
&&-\frac{1}{2}\frac{h}{a^{2}}b\left(-4a^{2}c^{4}-24a^{4}c^{2}+4a^{6}+b^{2}c^{4}+2a^{4}b^{2}\right)\nonumber \\
 & - & {m^2 \over \ell^6} \left[\frac{1}{2 he^{3z}}\left(e^{4\gamma-\beta-2\alpha}(v_{1}')^{2}+2e^{2\alpha-\beta}(v_{2}')^{2}+4e^{\beta}(v_{3}')^{2}\right) \right.\nonumber \\
 &  & +h e^{-3z}\left(2e^{-\beta}(v_{1}+v_{2})^{2}+e^{\beta-2\alpha}(v_{2}-v_{1}+2v_{3})^{2}+2e^{2\alpha+\beta-4\gamma}(2v_{3}-v_{2})^{2}\right)\nonumber \\
 &  &\left. \left.+  \left(3\tilde{z}'e^{-3\tilde{z}}\right)\left(4v_{3}(v_{1}+v_{2})+v_{2}(v_{2}-2v_{1})\right) \rule{0ex}{3ex} \right]\right\}
\eeq
where
\be a = e^{\alpha}, \qquad b = e^{\beta}, \qquad  c = e^{\gamma} \ . \ee
All of these fields $a$, $b$, $c$, $h$, $v_1$, $v_2$, $v_3$, $z$, and
$\tilde z$ can be viewed as being dimensionless. $r$ and $\ell$ have
dimension of length.

We can also use the reparametrization invariance of the radial variable to put  $h$ into a convenient form. We find it best to introduce the dimensionless radial variable $\tau$ using
\be h\,  dr = - \ell a^2 b c^4 \, d \tau \ . \label{r-tau}\ee
This eliminates $h$ as an effective degree of freedom aside from imposing the zero energy condition.

In order to apply the procedure of Borokhov and Gubser, we would like to organize this action into a form where
\be S = -\int d\tau\,  (T-U) \label{effTU}
\ee
where the potential $U$ can be expressed in terms of some
superpotential. This is accomplished by the procedure of integrating
out $K = (e^{- 3 \tilde z})'$ at the expense of introducing an
auxiliary parameter $q$ as we did in (\ref{Kdef}).

These manipulations  will bring the effective action into the form (\ref{effTU}), with
\beq
T & = & 2(\alpha')^{2}+12(\gamma')^{2}+4\alpha'\beta'+16\alpha'\gamma'+8\beta'\gamma'-\frac{9}{2}(z')^{2}\nonumber \\
& &- {m^2 \over \ell^6} \frac{1}{2a^{2}bc^{4}e^{3z}}\left(e^{4\gamma-\beta-2\alpha}(v_{1}')^{2}+2e^{2\alpha-\beta}(v_{2}')^{2}+4e^{\beta}(v_{3}')^{2}\right)
\label{eq:Action Pre-Elimination}\\
U & = & \frac{1}{2}b^{2}c^{4}\left(-4a^{2}c^{4}-24a^{4}c^{2}+4a^{6}+b^{2}c^{4}+2a^{4}b^{2}\right)\nonumber \\
& & +  {m^2 \over \ell^6} \frac{a^{2}bc^{4}}{e^{3z}}\left(2e^{-\beta}(v_{1}+v_{2})^{2}+e^{\beta-2\alpha}(v_{2}-v_{1}+2v_{3})^{2}+2e^{2\alpha+\beta-4\gamma}(2v_{3}-v_{2})^{2}\right)\nonumber \\
& & +  \frac{e^{-6z}}{2} \left(\frac{m^{2}}{\ell^{6}}\left(4v_{3}(v_{1}+v_{2})+v_{2}(v_{2}-2v_{1})\right)+q\right)^{2}
\eeq
for
\be
W  =  -bc^{2}\left(4a^{3}-2a^{2}b+4ac^{2}+bc^{2}\right) +  e^{-3z}\left(\frac{m^{2}}{\ell^{6}}\left(4v_{3}(v_{1}+v_{2})+v_{2}(v_{2}-2v_{1})\right)+q\right)
\label{Superpotential}
\ee

To identify the background solution, we set up the BPS equation (\ref{BPSeq}) for
\be \{ \phi^1, \phi^2, \phi^3, \phi^4, \phi^5, \phi^6, \phi^7\} = \{\alpha, \beta, \gamma, z, v_1, v_2, v_3\}  \ .  \ee
with $\tilde z$ is determined by the auxiliary condition
\be
3\tilde{z}'e^{6z-3\tilde{z}}=q+\frac{m^{2}}{\ell^{6}}
\left(4v_{3}(v_{1}+v_{2})+v_{2}(v_{2}-2v_{1})\right) \ . 
\label{ztilde} 
\ee
%

%



One can immediately confirm, for example, that flat space in eleven
dimensions,
\be a =b=c= (3 \tau)^{-1/6} \qquad v_i = 0 \ , \ee
is a solution, which can be presented in the standard form by
parametrizing
\be {3 \tau = \left({2 \ell \over r}\right)^{6}} \ .  \ee

Similarly, the BPS $A_8$ solution of  section \ref{sgA8} can be shown to solve the BG equations. The BPS equation  (\ref{BPSeq}) is invariant under reparametrization of coordinates as long as one suitably transforms
\be G^{ij} \rightarrow {d \tau \over d r} G^{ij} \ee
and one can confirm that the $A_8$ solution is indeed a solution of
(\ref{BPSeq}) in the $r$ coordinates.  The $r$ and $\tau$ coordinates are related by
\be \tau = - \int {dr \over \ell} \, {h \over  a^2 b c^4}= 
\frac{\ell^3}{3\left(r-\ell\right)^{3}}-\frac{\ell^2}{4\left(r-\ell\right)^{2}}+\frac{3\ell}{16\left(r-\ell \right)}
 +  \frac{\ell}{16\left(r+3\ell \right)}+\frac{1}{16}\log{\frac{r-\ell}{r+3 \ell}} \ . 
\label{COV}
\ee
In the dimensionless $\tau$ coordinates,  the kinetic term  the metric $G_{ij}$ take on a relatively simple form.

Expanding in inverse power of $\tau$ is equivalent to expanding near the tip $r=\ell$.  The warp factor (\ref{warpH}) for the warped $A_8/Z_k$ background can be written, as an expansion in $\tau^{-1}$, as
\be e^{3z} = q \tau + 
\left({m^2 \over \ell^6} {23 \over 5 \times 2^{12}} - q {\cal O}(\tau^0)\right) + {\cal O}(\tau^{-1/3}) \label{warpH0} \ . 
\ee
Comparing to the form of (\ref{warpH}), we find that 
\be q \ell^6 = {3 \over 2}  \pi^2 (l_p^{11})^6 k Q \label{Qq} \ee
where $q$ is the parameter appearing in the superpotential and $Q$ is the D2 charge normalized such that a single D2-brane has charge one.

We will take this solution at $q=0$ as our background
solution $\phi^i_0$  and expand
\be \phi^i = \phi^i_0 + \phi^i_1 \ee
and attempt to find the $\xi_i$ and $\phi^i_1$ which describe the back
reaction of an anti D2-brane to first order. 
We also allow $q$ to
shift at the same order as $\xi_i$ and $\phi_1^i$.

To proceed further, we need to take a closer look at the structure of
the matrix $N^j{}_i$ entering the Borokhov-Gubser equation (\ref{BG}). To this end, it is convenient to group the 7 fields
$\phi_1^i$ into subgroups which we might refer to as
\beq 
\phi_{geom} &=& \{ \phi_1^1, \phi_1^2, \phi_1^3 \} = \{\alpha, \beta, \gamma\} \cr
\phi_{z} & = & \{\phi_1^4 \} = \{ z \} \cr
\phi_{flux} &  = &  \{ \phi_1^5, \phi_1^6,  \phi_1^7 \} =\{v_1, v_2, v_3 \} 
\eeq

In this classification, $N^j()_i$ has a block structure which can be
summarized by
\be N^j{}_i = \left(\begin{array}{c|c|c} N_{geom} & & \\ \hline  & N_{z} & N_{zf} \\ \hline N_{gf} & & N_{flux} \end{array}\right)_i^j \ee

There are various blocks which have vanishing entries, which suggests
a strategy for the order of solving the Borokhov-Gubser equations. Specifically
\begin{enumerate}
\item We first solve for $\xi_4$ which is  decoupled and can be solved in closed form,
\item then we solve for $\xi_{flux}= \xi_{5,6,7}$ which takes $\xi_4$ as a source but is otherwise decoupled,
\item then, we solve for $\xi_{geom}=\xi_{1,2,3}$ which takes other $\xi$'s as sources but are otherwise closed,
\item then, $\phi^{geom}=\phi_1^{1,2,3}$ form a closed set of equations taking $\xi$'s as sources,
\item then, $\phi^{flux}=\phi_1^{5,6,7}$ can be solved taking $\phi^{geom}$ and $\xi$'s as sources,
\item and finally, $\phi_1^4$ can be computed. 
\end{enumerate}
Although these steps are straightforward in principle, the fact that
the intermediate step involves diagonalizing a coupled system of three
first order differential equation makes this exercise somewhat
formidable to execute in complete form. This is in contrast to earlier
instances such \cite{Bena:2009xk,Bena:2010gs,Giecold:2011gw} where the
mixing only involved two fields which were significantly easier to
diagonalize. Fortunately, this analysis is still tractable if we
restrict the scope of our study to explore the asymptotic behavior
near $r=\ell$ or large $\tau$. It will turn out that this
suffices for the conclusion we are after. Let us now proceed to
describe this analysis in more detail. It should be stressed that
aside from logistical challenges, nothing prevents us from attempting
to explore the solution to the full system of equations numerically. 

Let us follow this procedure step by step. 
\begin{enumerate}
\item The first step of solving for $\xi_4$ 
\be {d \xi_4 \over d \tau} = -N^4{}_4 \xi_4 = e^{-3 z} \left({m^2 \over \ell^6} \left(4v_{3}\left(v_{1}+v_{2}\right) +v_{2}\left(v_{2}-2v_{1} \right)\right)+q\right) \xi_4
\ee
which, using (\ref{ztilde}) can be written  to linear order in $q$ and $\xi_4$ as
\be {d \xi_4 \over d \tau} =  3 z' e^{3z} \xi_4
\ee
can be solved by
\be \xi_4 = X_4 e^{3z} \sim X_4 {m^2 \over l^6} {23 \over 5 \times 2^{12}} + {\cal O}(\tau^{-1/3}) \ee
where $X_4$ is the integration constant for the $\xi_4$ equation, and we used (\ref{warpH0}) in the last step.

\item Now we are ready to consider the equations for $\xi_{567}$ which can
be written as
\be
\left(\begin{array}{c}
\xi_{5}'\\
\xi_{6}'\\
\xi_{7}'
\end{array}\right)=\left(\begin{array}{ccc}
0 & -e^{2\beta+4\gamma} & e^{2\alpha+4\gamma}\\
-2e^{4\alpha+2\beta} & e^{2\beta+4\gamma} & e^{2\alpha+4\gamma}\\
4\epsilon^{4\alpha+2\beta} & 2e^{2\beta+4\gamma} & 0
\end{array}\right)\left(\begin{array}{c}
\xi_{5}\\
\xi_{6}\\
\xi_{7}
\end{array}\right)+\frac{2}{3}{m^2 \over \ell^6}X_{4}\left(\begin{array}{c}
v_{2}-2v_{3}\\
v_{1}-v_{2}-2v_{3}\\
-2(v_{1}+v_{2})
\end{array}\right)
\ee

Now, this is a rather cumbersome equation to solve in closed from, but in the large $\tau$ $r \rightarrow \ell$ limit, reduces to 
\be
\left(\begin{array}{c}
\xi_{5}'\\
\xi_{6}'\\
\xi_{7}'
\end{array}\right)=\frac{1}{3\tau}\left(\begin{array}{ccc}
0 & -1 & 1\\
-2 & 1 & 1\\
4 & 2 & 0
\end{array}\right)\left(\begin{array}{c}
\xi_{5}\\
\xi_{6}\\
\xi_{7}
\end{array}\right)+{m^2 \over \ell^6}\frac{X_{4}}{16(3\tau)^{\frac{2}{3}}}\left(\begin{array}{c}
\frac{1}{3}\\
-\frac{1}{2}\\
\frac{1}{6}
\end{array}\right)
\ee
These equations can be solved, introducing integration constants $X_{5,6,7}$
\beq
\xi_{5} & = & \frac{1}{2}(X_{6}-X_{7})\tau^{2/3}-\frac{1}{2}X_{5}\tau^{-1}-\frac{3^{1/3}}{48}{m^2 \over \ell^6}X_{4}\tau^{1/3}\nonumber \\
\xi_{6} & = & X_{7}\tau^{2/3}-\frac{1}{2}X_{5}\tau^{-1}+\frac{3^{1/3}}{32}{m^2 \over \ell^6}X_{4}\tau^{1/3}\\
\xi_{7} & = & X_{6}\tau^{2/3}+X_{5}\tau^{-1}-\frac{1}{32\times3^{2/3}}{m^2 \over \ell^6}X_{4}\tau^{1/3}\nonumber 
\eeq
\item Now we feed these $\xi_{4,5,6,7}$ into the $\xi_{1,2,3}$ equations
and follow the same steps.  Below we only indicate the terms which are
singular in the large $\tau$ limit.
\beq
\xi_{1} & = & \frac{1}{2}X_{2}\tau\nonumber \\
\xi_{2} & = & \frac{1}{4}X_{2}\tau-X_{3}\tau^{1/3}\nonumber \\
\xi_{3} & = & X_{2}\tau+X_{3}\tau^{1/3}
\eeq
\item Next, we compute $\phi_1^{1,2,3}$.
\begin{eqnarray}
\phi_1^{1} & = & {m^2 \over \ell^6}\frac{1}{32,768}X_{4}\tau^{2/3}-\frac{41}{3,072\times3^{2/3}}X_{6}\tau-\frac{1}{384\times3^{2/3}}X_{7}\tau \cr
&& -\frac{5}{2}(0  X_3 + Y_{1})\tau^{4/3}+\frac{X_{2}}{144}\tau^{2}\nonumber \\
\phi_1^{2} & = & {m^2 \over \ell^6}\frac{73}{98,304\times3^{1/3}}X_{4}\tau^{2/3}+\frac{(-8X_{6}+25X_{7})}{768\times3^{2/3}}\tau\cr
&& +\frac{1}{10}(3X_{3}+10Y_{1})\tau^{4/3}+\frac{X_{2}}{144}\tau^{2}\nonumber \\
\phi_1^{3} & = & -{m^2 \over \ell^6}\frac{25}{98,304\times3^{1/3}}X_{4}\tau^{2/3}+\frac{(31X_{6}-26X_{7})}{3,072\times3^{2/3}}\tau\cr
&&+\frac{1}{40}(-3X_{3}+40Y_{1})\tau^{4/3}+\frac{X_{2}}{144}\tau^{2}\nonumber\\ 
\label{PhiGeo}
\end{eqnarray}
We have included the $X_3$ term in $\phi_1^1$ whose coefficient is
accidentally zero. In generic linear combinations of $\phi_1^{123}$,
$X_3$ would appear in that order.

\item Now, computing the $\phi_1^{5,6,7}$  along similar lines, we find
\begin{eqnarray}
\phi_1^{5} & = & \frac{23}{983,040\times3^{2/3}}{m^2 \over \ell^6}X_{4}\tau^{1/3}-\frac{23}{163,840}(X_{6}-X_{7})\tau^{2/3}\nonumber \\
& & -  \frac{3^{1/3}}{640}(X_{3}-55Y_{1})\tau^{2/3}-2Y_{5}\tau+\frac{X_{2}}{2,304\times3^{2/3}}\tau^{4/3}\nonumber \\
\phi_1^{6} & = & -\frac{23}{1,310,720\times3^{2/3}}{m^2 \over \ell^6}X_{4}\tau^{1/3}-\frac{23}{163,840}(0  X_6+ X_{7})\tau^{2/3}\nonumber \\
& & -  \frac{(27X_{3}-310Y_{1})}{2,560\times3^{2/3}}\tau^{2/3}-Y_{5}\tau-\frac{X_{2}}{3,072\times3^{2/3}}\tau^{4/3}\nonumber \\
\phi_1^{7} & = & \frac{23}{7,864,320\times3^{2/3}}{m^2 \over \ell^6}X_{4}\tau^{1/3}\cr
&&  +  \frac{\left(704\times3^{1/3}X_{3}-21,120\times3^{1/3}Y_{1}-23X_{6}+ 0 X_7\right)}{327,680}\tau^{2/3}\nonumber \\
&& +Y_{5}\tau+\frac{X_{2}}{18,432\times3^{2/3}}\tau^{4/3}\nonumber \\
\label{PhiFlux}
\end{eqnarray}
The coefficients $Y_5$, $Y_6$, and $Y_7$ can be viewed as parameterizing the magnitude of self-dual 4-forms of which only one linear combination corresponds to the square normalizable 4-forms.

\item Finally, 
\beq
\phi_1^{4} & = & -\left(\frac{49\times 3^{2/3}}{1,656}X_{3}+\frac{23}{184,320}\frac{m^{2}}{l^{6}}X_{4} -\frac{5\times3^{1/3}}{4,608}\left(X_{6}-2X_{7}\right)-\frac{35\times3^{2/3}}{828}Y_{1}\right)\tau \nonumber \\
 &  &+ Y_{4}+0 Y_1 \tau^{4/3} -\frac{35}{9,936\times3^{1/3}} X_{2}\tau^{5/3}+\frac{5\times2^{12}l^{6}}{69m^{2}}q\tau
\label{Phi4}
\eeq

\end{enumerate}

At this stage, we have enumerated the terms in $\phi_1^i$ which are
singular in the $\tau \rightarrow \infty$ limit, corresponding to
probing the near core region $r \sim \ell$. These expressions should
be viewed as representing, for each $X_i$ and $Y_i$, the leading
singular $\tau$ dependence in the large $\tau$ limit. Strictly
speaking, these solutions depend on all 14 $X_i$'s and $Y_i$'s, except
that only those which are singular are presented.  We have also
included the dependence on $q$ which is also to be treated at the same
order.

Out of this 15 dimensional space of linearized solutions, we wish to
identify the particular deformation corresponding to adding a specific
amount of anti D2-branes. We will attempt to determine the unique
linear combination based on 1) the zero energy condition, 2)
consideration of the effect of the deformation on the Page charge 3)
the expected forces on D2-brane probe, and 4) the requirement to keep
the solution regular near $r=\ell$.

As have been the case in many of the earlier works on related systems,
it will turn out that satisfying all of the 4 requirements appears to
be impossible.

\begin{enumerate}

\item Let us first consider the zero-energy condition which in the $r \approx \ell$ limit simplifies for these linear solutions to 
\begin{equation}
0=\xi_{i}\frac{d\phi_{0}^{i}}{d\tau}=-\frac{7}{24}X_{2}\label{eq:Zero Energy}
\end{equation}
Thus we need $X_{2}=0$ to first order.
Next, let us consider the effects of deformations $X_i$'s and $Y_i$'s
on D2, D4, and D6 Page charges. It turns out that all of these do not
affect any of the D2, D4, and D6 Page charges. The only parameter
which affects the D2 Page charge is the parameter $q$. So, the linear
deformation interpretable as adding anti D2-brane must shift $q$.

\item Next, we consider the force on D2-brane probe which is expected for
adding an anti D2-brane. This can be computed using the standard DBI
analysis giving rise to, using (\ref{ztilde})
\begin{eqnarray}
F & = & F^{DBI} + F^{WZ}\nonumber \\
& = & -3z'e^{-3z} + 3\tilde z'e^{-3\tilde z} 
\end{eqnarray}
which for the first order deformation simplifies to
\be 
F = 
{1 \over H(\ell)^2}\frac{23}{184,320}\frac{m^{2}}{l^{6}} X_{4}  \label{force}
\ee
\item Next, from the form of (\ref{Qq}) and (\ref{force}), we also infer that for $-Q$ anti D2-branes, we should scale
\be
X_{4}  =  - 3\left(\frac{5\times2^{12}}{23 \frac{m^{2}}{l^{6}}}\right)^{2} q
= -2\left(\frac{15 g_{s}^{2}k^{2}}{46}\right)^{2} \frac{k^3Q}{M^{4}}
\label{Normalization}
\ee
\item Next, we examine the divergences in the $\phi_1^i$'s in the core region. Looking at terms diverging as $\tau^{4/3}$ in $\phi_1^{1,2,3}$, we infer that $Y_1$ and $X_3$ should be set to zero. This leaves terms diverging as $\tau$ in $\phi_1^{1,2,3}$. From this, we see that $X_6$ and $X_7$ should also be set to zero to prevent singularities in the core region.  So far, from looking at $\phi_1^{1,2,3}$ alone, we have set
\be  \{X_3, X_6, X_7, Y_1 \}\ee
to zero, in addition to $X_2$ which had to vanish because of the
zero energy condition. We will hold off on addressing $X_4$ for now
since that mode plays a special role in coupling to the D2-brane
probe.

Just from these constraints, we see that $\phi_1^{4,5,6}$ are also
severely constrained. Aside from $X_4$, the only remaining integration
constant is $Y_5$. $Y_5$ appears to be an interesting mode which we
will further discuss elsewhere. It corresponds to deformation by
non-normalizable self-dual 4 form, as can be seen as arising as a
$Y$-deformation, which is supersymmetry preserving.

The other constants 
\be \{X_1, X_5, Y_2, Y_3,  Y_6, Y_7 \}\ee
do not appear to induce divergent terms in the core region.

\end{enumerate}

The ultimate question whose answer we seek is whether one can deform
the solution to incorporate the back reaction of anti D2-branes while
preventing additional singularities from appearing.  It appears that
the answer, as was the case in many earlier attempts in related
systems, is ``no.'' In order to capture the tension of the anti
D2-brane in the warp factor, we have to turn on $X_4$. That $X_4$
turns on $\tau^{2/3}$ singularities in $\phi_1^{1,2,3}$ and $\tau^{1/3}$
singularities in $\phi_1^{5,6,7}$, and there are no remaining
adjustable integration constants one can turn on to cancel these
singularities without generating other singularities elsewhere.

This conclusion is not extremely surprising.  The narrative of how
the singularities in fluxes and the back reaction of the anti D2-brane
tension imposes conflicting constraints is identical to that which was
found in earlier analysis of similar constructions
\cite{Bena:2009xk,Bena:2010gs,Bena:2011hz,Bena:2011wh,Giecold:2011gw,Bena:2012bk}. One
novel feature in our analysis is the explicit absence of any smearing
of the anti-brane sources. But this does not appear to have much
effect on the conclusion.

This however raises a question concerning the fate of the
non-supersymmetric ground state anticipated to encapsulate the
features illustrated in figure \ref{figb}. In the consideration of the
meta-stable vacua, it was always possible for the state to destabilize
under some repulsive effect generated by the breaking of
supersymmetry. In the case of the Chern-Simons theories under
consideration, however, one expects there to be a competing
restorative component to balance the dynamically generated repulsive
force. Also, unlike in the Klebanov-Strassler case, there are no
alternative BPS supergravity solution for a given $N$, $l$, $k$, and
$b_\infty$ where $Q < 0$. So there are no supersymmetric vacua for the
non-supersymmetric state to decay into.

In the discussion section, we will offer our speculation concerning
the fate of $Q<0$ theories from the perspective of the gravity dual.

\section{Discussions}

In the earlier sections, we formulated and analyzed the construction
of supergravity solutions corresponding to a warped $A_8$ geometry
parametrized by $N$, $l$, $k$, and $b_\infty$. In the analysis, it
became clear that there is a parameter,
\be Q = N - {l(l-k) \over 2k} - {k \over 24}  \ee
which if positive, gives rise to a sensible warped supergravity
solution with an asymptotic anti-de-Sitter region in the core. At
$Q=0$, we also found that there is a sensible supergravity solution
describing the geometry in the core region. The question was whether
one could explore the backgrounds for $Q < 0$. We examined this as an
exercise in incorporating the gravitational back reaction of anti
D2-branes added to the $Q=0$ background and studied the effect of this
operation at linear order in shift in $Q$. What we found is that
sensible perturbation respecting regularity in the core region while
accounting for the physical features of the anti D2-branes could not
be found.

Implicit in this thinking is the notion that starting from $Q=0$
solution, making $Q$ positive corresponds to adding a D2-brane and
making $Q$ negative corresponds to adding an anti D2-brane. This the
perfectly sensible way in which things work in familiar context such as
flat space which we reviewed in section \ref{sec:flat}.

One potential fallacy is the assumption that the switch from branes to
anti-branes should happen at $Q=0$ also for the $A_8/Z_k$
background. A hint that something might be tricky here stems from the
curvature correction term $-k/24$ in the expression for $Q$. Strictly
speaking, a correction of this form should be considered as part of
$\alpha'$ correction to supergravity since we assume we are working in
the strong 't Hooft coupling limit which instructs us to take
$Q_2^{Maxwell}$ to be large which amounts to considering $x=N/k$ and
$y=l/k$ to be large. The only reason we have to take the $-k/24$ term
seriously was the work of \cite{Drukker:2010nc} which, on the field
theory side, computed the free energy precisely for arbitrary 't Hooft
coupling, not just its strong coupling asymptotics. This means that
even in the $A_8$ background with no D2 or D4 branes added, i.e.\ with
$N=l=0$, the geometry has some D2 brane charge due to curvature
effects, and it contributes negatively.

Generally, in string theory, singular BPS geometries are considered
physically allowed if the object sourcing them exists in the
theory. For example, large curvature singularity near the fundamental
string solution is considered an acceptable singularity because a
fundamental string is part of string theory. Moreover, negatively
charged negative tension objects do not appear in classical gravity,
but the ones which arise from curvature corrections of orbifolds and
orientifolds \cite{Sen:1996zq,Sethi:1998zk} are exceptions to this
rule. A situation similar to this was part of the repulson/enhancon
construction \cite{Johnson:1999qt}.

Let us for a moment take the point of view that the class of solutions
we found for $Q>0$ can be extrapolated by changing $Q$ not down to
$Q=0$, but rather down to $Q = -k/24$. The negative $Q$ solution
exhibits a repulson type singularity were a generic massive object
will feel a repulsive force
\cite{Behrndt:1995tr,Kallosh:1995yz,Cvetic:1995mx}. Some of the
features of the repulson dynamics were discussed in
\cite{Hashimoto:2010bq} but were not taken too seriously at the time
because it was believed that the repulson geometry itself should not
be taken seriously. It is the fact that the curvature correction
$Q=-k/24$ appears also in the field theory analysis which offers a
renewed motivation to take the repulson solution seriously, at least
for $Q > -k/24$.

It is natural to wonder if this repulson singularity is resolved by
the enhancon mechanism \cite{Johnson:1999qt}. As far as we could tell,
this is not the case. The only BPS probe we are able to find is the
D2-brane. Looking at the kinetic term of the radial motion of D2-brane
in this background, we did not find any locus interpretable as the
enhancon radius.

\begin{figure}
\centerline{\includegraphics{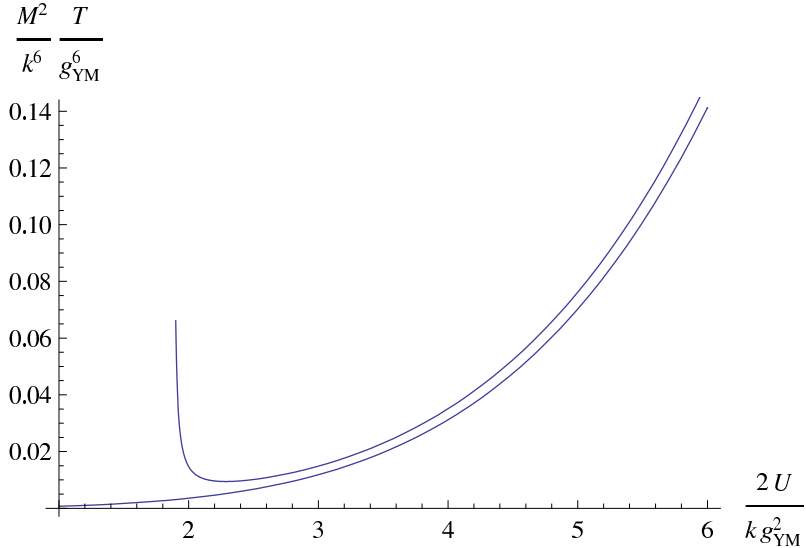}}
\caption{The potential experienced by an anti D2-brane in the $Q<0$ background inferred from the DBI action. This plot includes the extreme case $\epsilon = 0$ and a weakly repulsive case $\epsilon = 0.01$. \label{figc}}
\end{figure}

What we do find, as was reported originally in
\cite{Hashimoto:2010bq}, is that an anti D2-brane probe feels a
repulsive force near the core but is stabilized to sit a finite
radius. It is straightforward to compute the potential experienced by
the anti-D2-brane probe. It is simply
\be V = 2 T_2 H^{-1}(u) \label{antiD2V} \ee
where $u = {r / \ell} - 1 = 2 U / g_{YM}^2$, and so is stabilized where
\be F(u) = V'(u)   = 0 \ . \ee
The potential for $Q=0$ and small negative $Q$ is illustrated in
figure \ref{figc}. A little computation shows that as a function of
\be \epsilon = -{k Q \over M^2} \ee
the stabilization point scales as 
\be u \approx \epsilon^{1 /4} \ee
for small $\epsilon$. This was also noted in \cite{Hashimoto:2010bq}. 

If we take seriously the idea that it is $Q=-k/24$ and not $Q=0$ at
which we switch from subtracting branes to adding anti-branes as we
decrease $Q$, the narrative of the evolution of the supergravity
solution changes.

Let us take as given that we are always working with $Q \gg k$ so that
the supergravity description is good, and $Q \ll M^2 / k$ so that the
departure away from BPS solution can be considered parametrically
small. This of course implies that $M \gg k$. As we go from $Q$
positive to $Q$ negative in this parametrization, $Q$ negative and
small implies we have roughly $n$ anti D2-branes which we have added
to the $Q=-k/24$ solution for $1 \ll n/k \ll M$.

Out of these $n$ anti D2-branes, imagine adding them one by one to the $Q=-k/24$ background. The first one will feel the potential (\ref{antiD2V}) and settle at 
\be u \approx \epsilon^{1/4} \approx \sqrt{{k \over M}} \label{shellradius}\ee
where we are using the fact that 
\be Q = -{k \over 24} \ . \ee
For the purpose of making order of magnitude estimates, we have dropped the factor of $1/24$. 

As more anti D2-branes are added, each will stabilize at roughly the same radius, as anti D2-branes do not sense each other's presence, and we are neglecting the back reactions of these probe anti D2's for the time being.  It is natural to imagine these anti D2-branes forming a shell at the same radius as (\ref{shellradius}).

As even more anti D2's are added until we reach the total number  $n$,
one should not expect to get away with treating the anti D2's as a
probe. The configuration one might expect to find is that of the
$Q=-k/24$ geometry at the core, surrounded by a clump of anti
D2-branes which back react to build the full geometry. It should be
emphasized that these additional anti D2-branes do not make the
repulson singularity any stronger. The strength of the repulson always
corresponds to $Q=-k/24$, and the geometry is accompanied by a cloud
of anti D2-branes which floats in some distribution, balancing the
repulsion from the repulson and the stabilization due to the background
flux parametrized by $M$. It seems natural to imagine that all of the
$n$ anti D2-branes stabilize at a radius of the order
(\ref{shellradius}).

If this scenario is correct, one expects to find the supergravity
solution, along the lines of what we found using the linearized
perturbation around $Q=0$, but pushing $Q$ to be negative; i.e., we assume
that the solution is valid
for the radius outside the cloud of anti D2-branes. That some modes
develop a singularity near the core is no longer a problem because once
one hits the radius where the cloud of anti D2-branes is present, one
is expected to cross over into a different behavior of the
gravitational back reaction. In a sense, the cloud of anti D2 branes
shields the singularity implied by the perturbative analysis of section
\ref{sec:lin}.

In order to extract meaningful physical quantities characterizing the
dynamics of the $Q<0$ phase of this theory, however, one must first
come to grip with understanding how the anti D2-branes distribute
themselves in the region characterized by the radius
(\ref{shellradius}). Everything is happening at this very small
length-scale and it appears to be beyond the scope of supergravity to
settle this issue unambiguously.  One opportunistic scenario is that
the anti-D2-branes form a spherical shell of uniform density, and that
one can construct a full back reacted solution by joining the
$Q=-k/24$ solution on the inside and some generic $Q < 0$ solution on
the outside with the suitable matching condition at some appropriate
radius where a static solution can be shown to exist. It would be an
interesting exercise to see if such a solution can be constructed.

Regardless of this issue, our proposal is that there exists a
non-supersymmetric clump of anti D2-matter, stabilized by balancing a
repulsive force from curvature correction and an attractive force of
the background flux. This is a novel configuration of these objects in
string theory and may be relevant to characterizing the state
of other non-supersymmetric constructions.

Another consequence of this picture is the realization that the
singularities encountered in the perturbative analysis (section
\ref{sec:lin}) are a priori permissible because they can be
regularized by the $\alpha'$ corrections. One should add that it is
actually a bit of an over simplification. One can imagine that some of
these divergences can get regularized by the $\alpha'$ effects, but we
do not know if this is true of all singularities, nor how. In other
words, in assessing the field theory observables such as the
expectation values of some operators in the holographic language, one
would be interested in finding which normalizable modes are activated
in such a way that they are consistent with the boundary condition in
the core region. That boundary condition is precisely the information
encoded in the structure of the anti D2-brane clump in the core region
as well as the presumed sub-stringy physics regularizing the
repulson.  Understanding these issues brings the subject into the
treacherous terrain of the study of stable non-BPS configurations in
string theory and supergravity
\cite{Johnson:2000bm,Bertolini:2000jy,Bain:2000sa,Berglund:2000gt,Johnson:2001wm,Dimitriadis:2003ur}. All
of the quantitatively interesting information is encoded in the
stringy dynamics, and appears to be beyond the scope of a simple
space-time effective field theory analysis.

It should be emphasized, nonetheless, that the estimate of the mass
gap (\ref{Egap}) for $Q=0$ is a reliable prediction of the dual
gravity description. The scenario outlined above suggests that for
sufficiently small $\epsilon$, the scale of the gap will also make a
small change, but we are unable to infer the precise scaling without
making assumptions.

One may hope to make further progress on the field theory side. If the field theory side can access information at all orders in the 't
Hooft coupling, it will offer powerful insights into this
phenomenon. Unfortunately, the technique employed in \cite{Drukker:2010nc} is
not applicable for probing $Q < 0$ since it relies on
superconformal invariance as one of the key assumptions. By going to
$Q<0$, we are no longer able to rely on that feature.

Perhaps the most immediate task at hand is to explore the physics of
the anti-brane clump in a more controlled setting. In this paper, we
considered the regime $k \ll Q \ll M^2$ which forced $M$, the
parameter controlling the attractive forces involved in stabilizing
the anti D2-branes, to be large. This causes the size of the
anti-brane clump to be small. It would be interesting to see if
somehow one could make the strength $M$ of the attractive force small so
as to make the size of the clump large. In working with the supergravity
dual of a decoupled field theory system, it was a requirement that
$M^2/k$ be large in order to ensure that the 't Hooft coupling is
large. We can relax this requirement if we are going to study this
issue as a brane dynamic issue in the $A_8$ background without taking
the traditional $\alpha'\rightarrow 0$ limit where we ``drop the 1''
in the warp factor (\ref{warpH}).  With the ``1'' included, we can let
$M$ get close to the critical value $M^2\sim -2 Q k$ or $\epsilon =
kQ/M^2 \sim 1/2$ for $Q \sim -k/24$ and still have asymptotically
locally conical geometry.  Indeed, for $M^2 \sim -2 kQ$, one can show
that the anti D2-brane stabilization radius
\be r_* \sim -{16 kQ \over M^2 + 2 k Q} \sim {\epsilon \over 1 - 2 \epsilon}  \ee
can get arbitrarily large as $\epsilon \rightarrow 1/2$. However, the curvature of the potential at the minimum
\be V''(r_*) \approx {T Q \over g_s^4 k^5}  {(1 - 2 \epsilon)^8} \ee
is also getting smaller, indicating that the anti D2-branes would
likely spread out into a diffuse, as opposed to a thin, wall.  More details regarding this analysis can be found in appendix \ref{appB}.

\begin{figure}
\centerline{\includegraphics{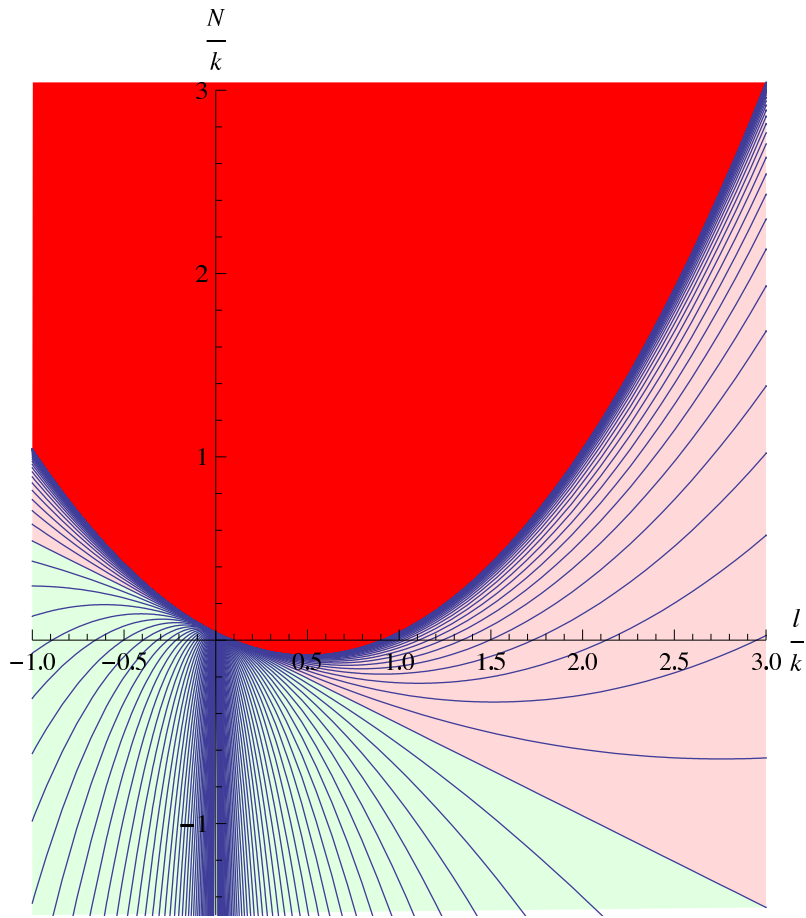}}
\caption{The phase diagram of the warped $A_8$ theory as a function of
$N/k$ and $l/k$. Here, we have set $b_\infty=1/2$ and $k$ is assumed
to be large. The red parabola indicates the region where $Q>0$ and the
theory flows to the superconformal fixed point of ABJM. Outside the red
parabola, we illustrate the contours of fixed $\epsilon$ in the range
$0 < \epsilon < \infty$ in logarithmic intervals. At $\epsilon = 1/2$,
the $Q_2^{Maxwell} = Q + M^2/2k$ changes sign, and we expect the
theory to transition into a new phase as $\epsilon$ crosses this line.
The supergravity approximation should be considered most reliable for
large values of $N/k$ and $l/k$ and close to the red parabola
corresponding to small values of $\epsilon$.
\label{figd}}
\end{figure}

It is also interesting to note that when $\epsilon > 1/2$, the Maxwell charge at infinity
\be Q^{Maxwell} = Q + {M^2 \over 2k} \ee
flips sign.  At this point, the stabilization radius $r_*$ no longer
exists for finite $r$.  This appears to suggest that the system
undergoes some kind of phase transition at $\epsilon = 1/2$. For the
sake of illustration, we have drawn the fixed $\epsilon$ contours for
$b_\infty = 1/2$ and for range of values $0 < \epsilon < \infty$ in
logarithmic scale in figure \ref{figd}. The plot is very similar to
figure 12 in \cite{Hashimoto:2010bq}.  The phase transition at
$\epsilon = 1/2$ is illustrated as the transition from the light red
to the light green region.

In closing, let us also comment on a potentially interesting
possibility of exploring the non-supersymmetric configuration
corresponding to having $Q \ll -k< 0$ but having small $M^2$ which we
treat as a perturbation. For $M^2=0$, the system should be described
in supergravity by the standard anti D2 solution in $A_8/Z_k$ with no
self-dual 4-form turned on. Turning on a small self-dual 4-form will
break supersymmetries incompatible with the anti D2-brane, and as
such, leads to a non-supersymmetric solution. This can be explored
either in the non-decoupled, i.e. for $H(r \rightarrow \infty) = 1$
solution, or the decoupled solution $H(r \rightarrow \infty) = 0$. It
is relatively straightforward to set up the Borokhov-Gubser type
analysis for this setup as well. The preliminary finding is that this
expansion is much better behaved. It should be noted from the outset,
that working in the regime $M^2 \ll -Q k$ is tantamount to working
with $\epsilon \gg 1/2$ and so is deep in the region which we believe
is in a different phase than the $\epsilon < 1/2$ region, as can be
seen illustrated in figure \ref{figd}.  Nonetheless, this is part of
the full landscape of possible parametric choices for these models and
may teach us something interesting about non-supersymmetric dynamics
of field theory and string theory.

\section*{Note Added}

While this paper was in its final stages of preparation, a paper
\cite{Giecold:2013pza} appeared which has significant overlap on the
analysis of the linearized supergravity equations.  Our findings
regarding the linearized analysis appear to be in complete agreement
with \cite{Giecold:2013pza}.

\section*{Acknowledgements}

AH would like to thank Shinji Hirano and Peter Ouyang for the
collaboration \cite{Hashimoto:2010bq} in which some very useful
techical notes were developed.  This work supported in part by the DOE
grant DE-FG02-95ER40896.

\appendix

\section{D0 probe in $Q=0$ background}

In this appendix, we will briefly describe another brane probe one can consider for the $Q=0$ background which exhibits interesting behavior.

Consider  a probe D0-brane. One can compute the potential experienced by this probe simply using the DBI action
\be V(x) = {1 \over 2 \pi l_s g_s} e^{-\phi} \sqrt{g_{00}^{IIA}} = {1 \over 2 \pi l_s g_s} H(x)^{-1/2} b(x)^{-1} \ee
where $H(x)$ and $b(x)$ are given in (\ref{warpH}) and (\ref{habc}), and
\be r = \ell x \ee 

\begin{figure}
\centerline{\includegraphics{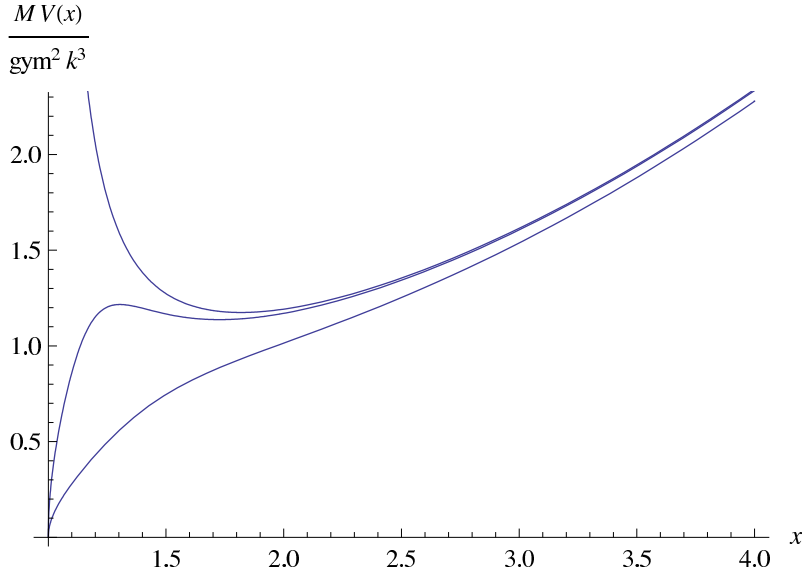}}
\caption{The potential of the D0-brane probe for $Q>0$ background inferred from the DBI action. At $Q=0$, the D0 brane is stabilized at a finite radius. For small and positive $Q$, the minimum at finite radius becomes metastable. As $Q$ is increased, the metastable minimum disappears and the D0 is attracted toward the core region at $r=\ell$. 
\label{fige}}
\end{figure}

For $Q\gg 0$, the D0 probe action is attractive for all $x$. However, for $Q=0$, there is a repulsive component to the potential, giving rise to a  potential illustrated in figure \ref{fige}.

When small positive $Q$ is turned on, this potential exhibits a metastable minimum until $Q$ reaches a critical value, at which point the metastable minimum goes away. 

In the perspective of gauge gravity duality, this is an object which
behaves as a localized magnetic flux which is stable and finite in
size. In some respects, this is a IIA version of the axion string
discussed in \cite{Gubser:2004qj}.  It might be interesting to further
explore the physics of this object.

\section{Anti D2-brane probe in a repulson background \label{appB}}

In this appendix, we provide the details of the anti D2 brane probe analysis in the repulson background. 
The anti D2 action takes the form
\be V(r) = T e^{-\phi} \sqrt{\det g_{\mu \nu}} = -T H^{-1}(r) \ee
for $H$ given in (\ref{warpH}), except that here, we also  include the additional ``1.''

The stable radius $r_*$ for the anti D2-brane probe, given by 
\be V'(r_*) =0 \ee
is equivalent to the condition that
\be H'(r_*) = 0 \ee
and so is insensitive to whether or not we include the ``1.'' 
Substituting (\ref{warpH}), we find that this condition reads
\be 
0  = 
(\ell-r_*)^4 \left(123 \ell^3+121 r_* \ell^2+33 r_*^2 \ell+3 r_*^3\right)-6 \epsilon
    (\ell+r_*)^3 (3 \ell+r_*)^4
\ee
which can be used to solve for $r_*$ in terms of $\epsilon$, but it is just as convenient to solve for $\epsilon$ in terms of $r_*$. Then, we find
\be \epsilon = 
\frac{(\ell-r_*)^4 \left(123 \ell^3+121 r_* \ell^2+33 r_*^2 \ell+3 r_*^3\right)}{6
   (\ell+r_*)^3 (3 \ell+r_*)^4} = {1 \over 2} - {4 \ell \over r_*} + {\cal O}\left(\left({\ell / r_*}\right)^2\right) \ee
Now, consider $V''(r_*)$. This expression depends critically on the ``1'' in the harmonic function. More specifically, this takes the form
\be V''(r=r_*) \sim {T  \over H(r=\infty)^2} {\ell^8 \over r_*^8}  \sim {T  \over H(r=\infty)^2} (1 - 2 \epsilon)^8 \ee
where we have only indicated the scaling with respect to $r_*$ for
large $r_*$ as the expression is somewhat complicated. Nonetheless,
the important point is that for $r_* \gg \ell$, i.e. for $\epsilon
\lesssim 1/2$, it goes to zero very rapidly.

\begin{figure}
\centerline{\includegraphics[width=3.45in]{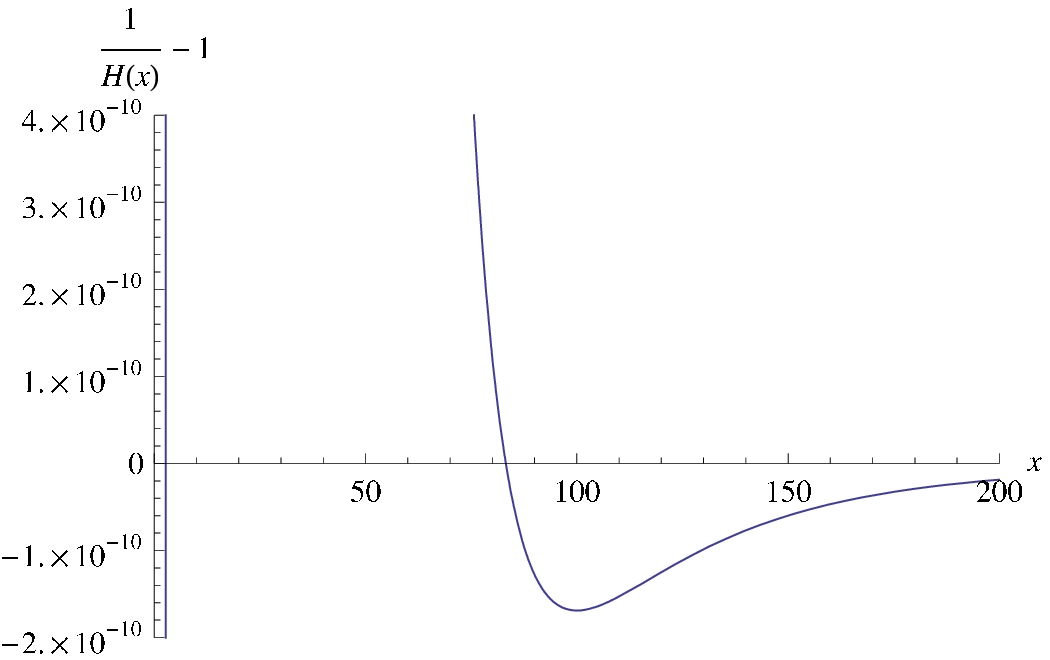}\qquad \includegraphics[width=3in]{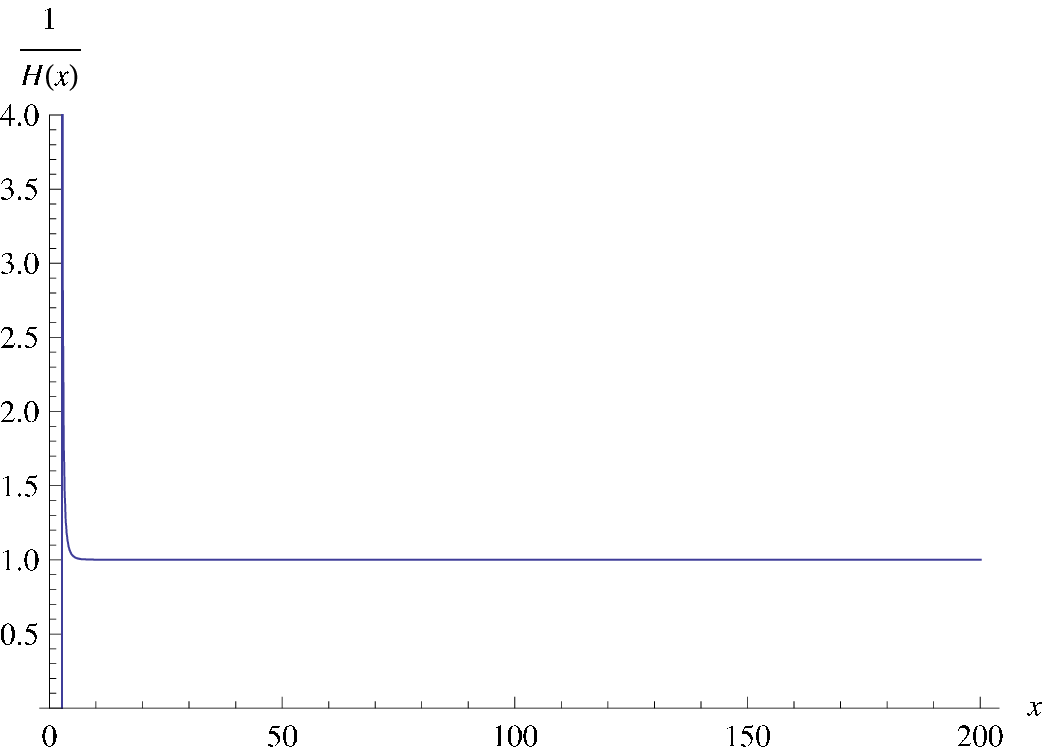}}
\caption{Anti D2 potential $H(x)^{-1}$ in the $Q = -k/24$ repulson background for the harmonic function $H(x)$ including the ``$1$.'' Here, we have set $g_s = k^{-1}$ and also set $x_*=100$, which  sets $M^2 \approx - 2 k Q $. The repulson radius is roughly at $x=2.69$. 
\label{figf}}
\end{figure}

It may be useful to illustrate the anti D2-brane potential in the case $r_* \gg \ell$  more explicitly. In terms of
\be x = {r \over \ell}, \qquad x_* = {r _* \over \ell} \ee
and for 
\be Q = - {k \over 24} \ee
the harmonic function $H$ can be written as
\beq
H(x) & =& 1 + {\pi^2 \over 60 g_s^4 k^4} 
\left[  -\frac{20 \left(3 x^3-3 x^2-11 x+27\right)}{(x-1)^3 (x+3)}-15
   \log \left(\frac{x-1}{x+3}\right)\right. \cr
&& \left. + \frac{768  (x_*+1)^3
   (x_*+3)^4}{ (x_*-1)^4 \left(3 x_*^3+33 x_*^2+121 x_*+123\right)}
{\left(3 x^2+26 x+63\right) \over (x+1)^2 (x+3)^5}
\right] 
\eeq
Note the presence of ``$1$'' in $H(x)$. A convenient choice to illustrate the possibility of separating $r_*$ from the repulson radius is to chose $g_s = k^{-1}$ and $x_*=100$. With this choice, the potential $V(x) \propto H(x)^{-1}$ has the form illustrated in figure \ref{figf}. We have specifically included the same plot in two different scaling of the axes in order to highlight the features in very large and very small scales. 

Suppose the idea that the BPS solution is reliable, at least for large
radius, down to $Q = -k/24$ with the understanding that string
dynamics self corrects the geometry in the $r < r_*$ region. What this
example illustrates is that one should anticipate string corrections
to impact regions all the way up to $r_* = 100 \ell$ which is
significantly further out than where one would have expected the
corrections based merely on the estimate of the curvature which is
concentrated in $r \approx \ell$ region. If correct, this would be a novel mechanism to induce larger corrections to gravity than what one would naively expect in effective field theory considerations.

\bibliography{nonsusy}\bibliographystyle{utphys}

\end{document}